\documentclass[12pt,a4paper]{article}

\usepackage{amsmath,amsfonts,amssymb}
\usepackage{hyperref}
\usepackage{graphicx}
\usepackage[font=small,labelfont=bf]{caption}
\usepackage[font=footnotesize,labelfont= small]{subcaption}
\usepackage{color}
\usepackage{epsfig}
\usepackage{psfrag}
\usepackage{slashed}
\usepackage{ulem}
\usepackage{cite}
\usepackage{url}
\usepackage{braket}
\usepackage[lmargin=2.5cm,rmargin=2.5cm,bmargin=3cm,tmargin=2cm]{geometry}
\usepackage{mathabx}
\usepackage{float}

\usepackage{color}
\newcommand \be{\begin{eqnarray}}
\newcommand \ee{\end{eqnarray}}

\makeatletter\@addtoreset{equation}{section}\makeatother

\DeclareMathOperator{\Tr}{Tr}

\DeclareMathOperator{\Ai}{Ai}

\DeclareMathOperator{\sh}{sh}
\DeclareMathOperator{\ch}{ch}
\DeclareMathOperator{\thh}{th}
\DeclareMathOperator{\Pf}{Pf}

\def\bC{\mathbb{C}}
\def\bD{\mathbb{D}}

\def\bZ{\mathbb{Z}}

\newcommand{\beq}{\begin{equation}}
\newcommand{\eeq}{\end{equation}}
\newcommand{\bal}{\begin{equation}\begin{aligned}}
\newcommand{\eal}{\end{aligned}\end{equation}}

\newcommand{\eqn}[1]{(\ref{#1})}

\newcommand{\address}[1]{\vbox{\center\em#1}}
\renewcommand{\title}[1]{\vbox{\center\huge{#1}}\vspace{5mm}}

\newcommand{\cG}{{\mathcal G}}

\newcommand{\cK}{{\mathcal K}}

\newcommand{\cN}{{\mathcal N}}

\newcommand{\cR}{{\mathcal R}}
\newcommand{\cO}{{\mathcal O}}

\newcommand{\cW}{{\mathcal W}}

\begin{document}

\begin{titlepage}
\begin{center}
\phantom{}

\vspace{15mm}

\title{Partition functions of 3d $\hat D$-quivers and their mirror duals from 1d free fermions}
\vspace{5mm}

\renewcommand{\thefootnote}{$\alph{footnote}$}

Benjamin Assel\footnote{\href{mailto:benjamin.assel@gmail.com}{\tt benjamin.assel@gmail.com}},
Nadav Drukker\footnote{\href{mailto:nadav.drukker@gmail.com}{\tt nadav.drukker@gmail.com}}
and Jan Felix\footnote{\href{mailto:jan.felix@kcl.ac.uk}{\tt jan.felix@kcl.ac.uk}}
\vskip 5mm 
\address{
Department of Mathematics, King's College London \\
The Strand, WC2R 2LS, London, UK}

\renewcommand{\thefootnote}{\arabic{footnote}}
\setcounter{footnote}{0}

\end{center}

\vskip5mm 

\abstract{
\normalsize 
\noindent 
We study the matrix models calculating the sphere partition functions of 
3d gauge theories with $\cN=4$ supersymmetry and a quiver structure of 
a $\hat D$ Dynkin diagram (where each node is a unitary gauge group). 
As in the case of necklace ($\hat A$) quivers, 
we can map the problem to that of free fermion quantum mechanics whose 
complicated Hamiltonian we find explicitly. Many of these theories are conjectured to be 
dual under mirror symmetry to certain unitary linear quivers with extra $Sp$ nodes 
or antisymmetric hypermultiplets. We show that the free fermion formulations of 
such mirror pairs are related by a linear symplectic transformation.

We then study the large~$N$ expansion of the partition function, which as in the case 
of the $\hat A$-quivers is given to all orders in $1/N$ by an Airy function. We simplify the 
algorithm to calculate the numerical coefficients appearing in the Airy function and evaluate 
them for a wide class of $\hat D$-quiver theories.
}

\end{titlepage}

\section{Introduction}
\label{sec:intro}

Three-dimensional gauge theories with $\cN=4$ supersymmetry have an
$SU(2)_C \times SU(2)_H$ $R$-symmetry group, with $SU(2)_C$ acting 
on the Coulomb branch moduli and $SU(2)_H$ acting on the Higgs branch moduli.
A large class of three-dimensional $\cN=4$ gauge theories flow to interacting fixed 
points in the infrared limit. These theories possess a remarkable duality known as 
mirror symmetry \cite{Intriligator1996}, which is the statement that pairs of UV gauge 
theories flow to the same infrared fixed point with the action of $SU(2)_C$ and 
$SU(2)_H$ exchanged. In particular the classical (protected) Higgs branch of 
one theory matches the quantum-corrected Coulomb branch of the other theory 
\cite{Boer1997a, Boer1997}. The exact results obtained from the technique of 
supersymmetric localization provide a privileged testing ground for mirror symmetry. 
In particular the full partition function of 3d $\cN=4$ gauge theories defined on a 
three-sphere reduces to a relatively simple matrix model \cite{Kapustin2010}. 
This matrix model is independent of running coupling constants and therefore 
computes the three-sphere partition function of the infrared fixed point. Mirror 
symmetry was successfully tested by matching the matrix models of pairs of 
mirror-dual theories in \cite{Kapustin2010a, Dey2013, Assel2014}. More tests 
of mirror symmetry by matching the Coulomb branch and Higgs branch Hilbert 
series of dual theories were achieved in 
\cite{Cremonesi:2013lqa, Cremonesi:2014vla, Cremonesi:2014uva, Dey2014a}.

The most studied $\cN=4$ gauge theories subject to mirror symmetry are infrared fixed points of quiver 
theories of type $\hat A, \hat D$ or $\hat E$, referring 
to the shape of the quiver
as an 
extended Dynkin diagram and likewise to the 
orbifold singularity in their M-theory realization \cite{Porrati:1996xi, Dey2012}.
It was found in \cite{Drukker2015} that the mapping between matrix models of 
mirror-dual theories of type $\hat A$ can be expressed as a very simple canonical 
transformation, exchanging position and momentum, in the 1d free fermion formalism 
developed in \cite{Marino2012}. 

The free fermion formalism arises from the observation that the matrix model 
computing the three-sphere partition function of $\cN \ge 3$ Chern-Simons-Matter 
quiver theories can be re-expressed as the partition function of a gas of non-interacting 
fermions in one dimension with a non-trivial Hamiltonian. This formalism allowed the use of powerful 
techniques from quantum and statistical mechanics 
to solve the matrix models as an $\hbar$ expansion, $\hbar$ being related to the 
Chern-Simons levels of the 3d theory. It was found in \cite{Marino2012} that the perturbative 
part of the partition function at large~$N$ takes the form of an Airy function. 

This Airy function behavior was first found for the particular case of ABJM theory 
\cite{Aharony2008}, in \cite{Fuji2011}, based on its original large~$N$ perturbative 
solution \cite{Drukker:2010nc}. Non perturbative corrections have been studied 
intensively for ABJM \cite{Drukker:2011zy,Hatsuda2012, Putrov2012, Hatsuda2013, 
Calvo2013,Hatsuda2013a, Hatsuda2014a,Kallen2013,Kallen2014}, 
ABJ \cite{Matsumoto2014, Honda2014a }, and some other examples of $\hat A$-quiver theories \cite{Hatsuda2014, Moriyama2014b}. 
Recently, the grand partition function of ABJ(M) was determined exactly 
\cite{Codesido2014, Grassi2014}, in perfect agreement with numerical results \cite{Wang2014}.

\begin{figure}[ht]
\centering
\epsfig{file=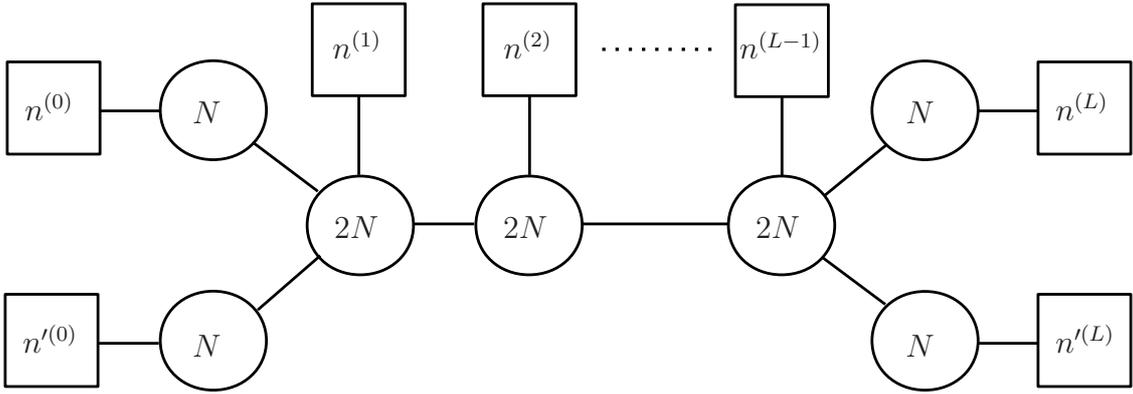,width=15cm
\psfrag{2N}{$2N$}
\psfrag{Nf0}{$n^{(0)}$}
\psfrag{Nf0t}{$ n^{\prime(0)}$}
\psfrag{N0}{$N$}
\psfrag{N0t}{$N$}
\psfrag{Nn2}{$N$}
\psfrag{Nn2t}{$N$}
\psfrag{Nf1}{$n^{(1)}$}
\psfrag{Nf2}{$n^{(2)}$}
\psfrag{Nfn3}{$\!\!\!n^{(L-1)}$}
\psfrag{Nfn2}{$n^{(L)}$}
\psfrag{Nfn2t}{$n^{\prime(L)}$}
}
\caption{\label{fig:Dgen}%
The general $\hat D_{L+2}$-quiver with arbitrary fundamental matter.}
\end{figure}

\begin{figure}[ht]
\centering
\begin{subfigure}{7cm}
\centering
\epsfig{file=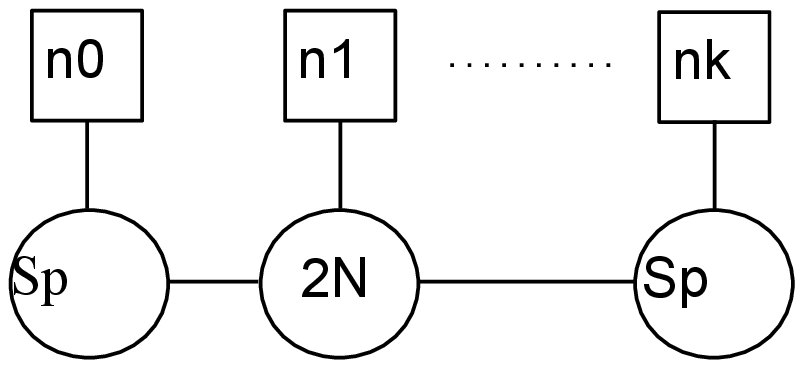, width=7cm
\psfrag{Sp}{\small$Sp(2N)$}
\psfrag{2N}{\!\!\!\small$U(2N)$}
\psfrag{n0}{$n^{(0)}$}
\psfrag{n1}{$n^{(1)}$}
\psfrag{nk}{$n^{(L)}$}
}
\subcaption{}
\label{fig:SpSp}
\end{subfigure}
\qquad\quad
\begin{subfigure}{7.5cm}
\centering
\epsfig{file=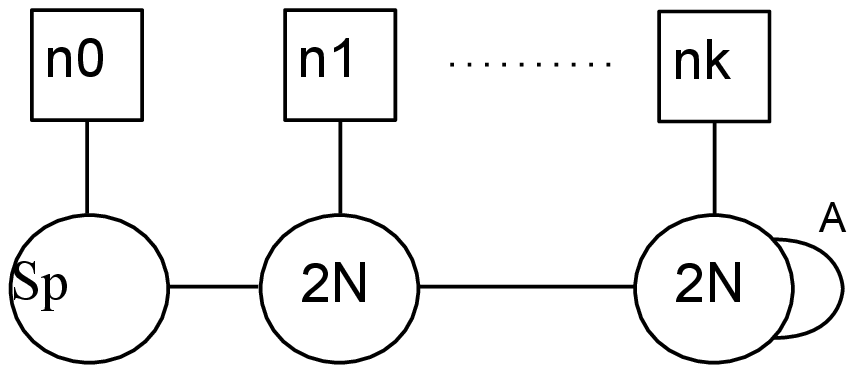, width=7.5cm
\psfrag{Sp}{\small$Sp(2N)$}
\psfrag{2N}{\!\!\!\small$U(2N)$}
\psfrag{n0}{$n^{(0)}$}
\psfrag{n1}{$n^{(1)}$}
\psfrag{nk}{$n^{(L)}$}
\psfrag{A}{$A$}
}
\subcaption{}
\label{fig:SpA}
\end{subfigure}
\caption{\label{fig:Spgenfund}%
Examples of linear quivers that are mirror dual to $\hat D$-quivers}
\end{figure}

In this paper we find the free fermion formalism associated to infrared fixed points of $\cN=4$
Yang-Mills quiver theories of type $\hat D$ 
and similar Chern-Simons-matter theories.%
\footnote{Chern-Simons terms generically lead to only $\cN=3$ 
supersymmetry, and we mostly set them to zero after section~\ref{sec:fermi}.}
A $\hat D_{L+2}$-quiver is described by a linear chain 
of $(L-1)$ $U(2N)$ gauge nodes, connected by bifundamental hypermultiplets, 
with pairs of $U(N)$ nodes attached at both ends of the $U(2N)$ chain, as 
shown in figure~\ref{fig:Dgen}. In addition each node may couple to an arbitrary 
number of fundamental hypermultiplets, indicated by the boxes 
in the diagram.
Contrarily to type $\hat A$ theories, mirror symmetry does not relate pairs of type 
$\hat D$ theories. Instead, $\hat D$-quivers are mapped under mirror symmetry to 
linear quivers with $U(2N)$ gauge nodes and extra symplectic gauge nodes $Sp(2N)$ 
or antisymmetric hypermultiplets at both ends of the quiver chain 
\cite{Hanany:1999sj, Gaiotto2009, Dey2012, Dey2014}, that we call generically 
{\it linear quivers}.%
\footnote{There are many other types of linear quivers, which we do not study in this paper. 
We use this name to refer to the quivers as in figure~\ref{fig:Spgenfund} and hope that this 
will not cause confusion.}

Roughly speaking, the length of the linear quiver mirror depends on the number of fundamental 
hypermultiplets coupling to the $U(2N)$ nodes and the structure of the end of the linear 
quiver depends on the number of fundamental hypermultiplets 
on the $U(N)$ nodes. 
When the number of fundamental hypermultiplets 
on the two $U(N)$ nodes on one side of 
the $\hat D$-quiver are equal, the mirror theory has a terminating $Sp(2N)$
node. When the numbers of fundamental hypermultiplets differ by one, the mirror theory has a 
terminating antisymmetric hypermultiplet. For instance if the numbers of 
$U(N)$ fundamental hypermultiplets are given by $n^{(0)} = n^{\prime(0)}= 0$ and 
$ n^{(L)} = n^{\prime(L)}=1$ (in the notations of figure~\ref{fig:Dgen}), then the mirror 
theory has $Sp(2N)$ nodes at each end as in figure~\ref{fig:SpSp}, while 
$n^{(0)} = n^{\prime(0)}= n^{(L)} = 0$ and $ n^{\prime(L)}=1$ leads to a 
mirror linear quiver with a terminating $Sp(2N)$ node at one end and an antisymmetric 
hypermultiplet at the other end as in figure~\ref{fig:SpA}.%
\footnote{Note that the parameters $n^{(a)}$ in figures~\ref{fig:Dgen} 
and~\ref{fig:SpSp}, \ref{fig:SpA} do not get mapped to each other under mirror symmetry. 
The actual mirror map is more involved, see section~\ref{sec:mirror}.}

When the numbers of fundamental hypermultipets on the two $U(N)$ 
nodes differ by two or more, we do not know what is the precise mirror dual theory. 
Naive considerations on the Hanany-Witten 
IIB brane setup realizing these quivers \cite{Hanany1997, Gaiotto2009}, lead to `bad' mirror 
dual linear quivers \cite{Dey2014a}. `Bad' quivers are quivers whose gauge group cannot be
completely higgsed and whose three-sphere partition function is divergent.
The formal manipulations leading to a Fermi gas description presented in section 
\ref{sec:fermi} can be done for any quiver, however when we study mirror symmetry, 
we focus only on duality between `good' quivers.

Our starting point is the matrix model 
computing the three-sphere partition function of the quiver theories. 
This matrix model is expressed as 
an integral over the Cartan subalgebra of the gauge group and the integrand is a 
product of classical contributions and one-loop contributions in the (exact) saddle 
point analysis of supersymmetric localization~\cite{Kapustin2010}
\beq
Z = \frac{1}{|\cW|} \int_{\rm Cartan} d\lambda \,Z_\text{class}\cdot Z_{\text{vec}}\cdot Z_{\text{hyp}}\,,
\label{3dZ}
\eeq
where $|\cW|$ is the order of the Weyl group.
The classical contribution depends on 
Fayet-Iliopoulos (FI) and Chern-Simons (CS) parameters. These are given by 
\beq
Z^{\text{FI}}_\text{class}=e^{2 \pi i\, \zeta \Tr \lambda}\,, 
\qquad 
Z^{\text{CS}}_\text{class} = e^{ \pi i\, k \Tr \lambda^2}\,, 
\label{Zclass}
\eeq
with $\zeta$ the FI parameter, $k$ the Chern-Simons level and Tr the trace in 
the fundamental representation. 
The one-loop contributions of the $\cN=4$ vector multiplet and hypermultiplet in a 
representation $\cR$ are
\beq
Z_{\text{vec}}= \prod_{\alpha>0} 4 \sinh^2(\pi \, \alpha\cdot\lambda)\,, 
\qquad
Z_{\text{hyp}} =\prod_{w\in {\cR}} \frac{1}{2\cosh(\pi ( w\cdot \lambda + m))}\,,
\eeq
where $\alpha$ runs over the positive roots of the Lie algebra and $w$ over the 
weights of $\cR$. $m$ is a real mass parameter for the hypermultiplet. We provide 
in appendix~\ref{sec:matrixmodels} the explicit matrix factors relevant to the 
$\hat D$-quivers and their mirror linear quivers. 

By manipulating the matrix models of $\hat D$-quivers we are able to 
re-express it in the form
\beq
Z_{\hat D} = \frac{1}{N!}\sum_{\sigma \in S_N} \frac{(-1)^\sigma}{2^{n_\sigma}} 
\int \prod_{i=1}^N d\lambda_i \,\prod_{i=1}^N \rho(\lambda_i, \lambda_{\sigma(i)})\,,
\label{ZD}
\eeq
where $\rho$ depends on the content of the theory and $n_\sigma$ is the number 
of cycles in the permutation $\sigma$. This expression differs 
from the analogue expression for $\hat A$ quivers by the presence of the factor 
$1/2^{n_\sigma}$. Defining the density operator $\hat\rho$ by 
$\bra{\lambda}\hat\rho \ket{\lambda'} = \rho(\lambda, \lambda')$,
we are able to recast the partition function 
of theories with vanishing mass and FI parameters into
\beq
Z_{\hat D}= \frac{1}{N!}\sum_{\sigma \in S_N} (-1)^\sigma 
\int \prod_{i=1}^N d\lambda_i \,\prod_{i=1}^N 
\bra{\lambda_i} \hat\rho \bigg( \frac{1 + \hat R}{2} \bigg) \ket{\lambda_{\sigma(i)}}\,,
\label{ZD1d}
\eeq
where $\hat R$ is the reflection operator $\hat R \ket{\lambda} = \ket{-\lambda}$. 
This can be interpreted as the partition function of $N$ non-interacting 
fermions on a {\it half line} with Neumann boundary condition at the origin, 
with a Hamiltonian $\hat H = - \log \hat\rho$.
We also find that $Z_{\hat D}$ can alternatively be recast into the partition function of $N$ 
non-interacting fermions on a half line with Dirichlet boundary condition at the origin, which 
translates into having the projection $\frac{1 - \hat R}{2}$ instead of $\frac{1 + \hat R}{2}$
in \eqref{ZD1d}. This freedom exists because, 
as we show in appendix~\ref{sec:degenspectrum}, 
the density operators associated with $\hat D$ 
quivers have the remarkable property that their spectrum is pairwise degenerate between even
and odd states on the line. 

Similarly we find that the matrix models associated to linear quivers can be set in 
the same form \eqref{ZD} or \eqref{ZD1d}, with the same property that the density 
operator has pairwise degenerate spectrum. 
The derivation of the density operators associated to the $\hat D$-quivers and their mirror-dual 
linear quivers are presented in section~\ref{sec:fermi}.

Having found the density operators for $\hat D$-quivers and linear quivers, 
we observe in section~\ref{sec:mirror} that density 
operators of mirror theories are the mapped by the transformation on position and 
momentum operators 
\bal
\label{mirrortransintro}
p \rightarrow q, \qquad q \rightarrow - p\,,
\eal 
as was first found in \cite{Drukker2015} for type $\hat A$ quivers.%
\footnote{Mirror symmetry for type 
$\hat A$-quivers extends to an $SL(2,\bZ)$ group of dualities 
\cite{Gaiotto2009, Assel2014}. We are not aware of a similar extension for 
$\hat D$-quivers and consequently our discussion concerns only the mirror 
transformation corresponding to the so-called $S$-duality.}

With a free fermion formalism, we are 
in a position to evaluate the {\it perturbative part} of the 
large~$N$ expansion of the $S^3$ partition function of $\hat D$-quivers, 
as was done for $\hat A$-quiver theories in \cite{Moriyama2014}. 
We present the analysis in section~\ref{sec:partition} and derive 
explicit expressions for theories with 
vanishing masses and FI parameters. This is conveniently described 
from the grand potential $J(\mu)$,
which is the logarithm of the grand canonical partition function
\beq
\Xi(z) = 1 + \sum_{N=1}^\infty Z(N) z^N = e^{J(\mu)}, \qquad z= e^\mu\,.
\label{ZToJ}
\eeq
Using {\it phase space} techniques, the grand potential evaluates to
\beq
\label{JmuDecomp}
J(\mu) = \frac{C}{3} \mu^3 + B \mu + A + \cO (e^{-\alpha \mu})\,, 
\qquad \alpha>0\,,
\eeq
which after inverting the relation \eqref{ZToJ} leads to
\beq
\label{ZNAiryintro}
Z(N) = C^{-\frac{1}{3} } e^A \Ai\left[C^{-\frac{1}{3}}(N -B ) \right] 
+ Z_{\text{np}}(N)\,, 
\eeq
where Ai is the Airy function and $ Z_\text{np}(N)$ denotes non-perturbative, 
exponentially suppressed contributions at large $N$. 
For quiver theories without CS terms and without mass or FI deformations, 
we are able to find the exact values of the coefficients $C$ and
$B$, but not $A$, nor the non-perturbative corrections.
The method we develop is simplified compared to earlier literature.

The fact that the result for $\hat D$-quivers is similar to that of $\hat A$-quivers 
is consistent with the localization calculation of supergravity in $AdS_4$ 
\cite{Dabholkar:2014wpa}. That suggests a universal answer for all conformal 3d theories 
with enough supersymmetry and an M-theory dual, which includes both the 
$\hat A$ and $\hat D$-quivers.

The free fermion formalism 
for 3d $\cN=4$ 
theories with no $U(2N)$ and a single $Sp(2N)$ or $SO(2N)$ 
node was studied previously in \cite{Mezei2014}. We compare our formalism 
to theirs in appendix~\ref{sec:prevfermi}.%
\footnote{We point out a mistake in their next-to-leading contribution to the grand potential.}

We conclude with a discussion of some open questions and possible 
extensions of our work in section~\ref{sec:discussion}. We also include a simple 
holographic test of our results. 

During the course of the work we learned that similar questions were studied by 
Moriyama and Nosaka, whose results are published concurrently \cite{MN}.

\paragraph{Notations:} 
We consider quivers with nodes of rank $N$ or $2N$ and we use the indices 
$i, j$ when the label runs over $1,\cdots, N$, and $I, J$ when the label runs 
over $1, \cdots, 2N$. 
Moreover we define
\beq
\sh x \equiv 2 \sinh \pi x\,,
\qquad
\ch x \equiv 2 \cosh \pi x\,,
\qquad
\thh x \equiv \frac{\sh x}{\ch x} = \tanh \pi x\,.
\eeq

\section{$\hat D$-quivers and linear quivers as free 1d fermions}
\label{sec:fermi}

In this section we show how to re-express the $S^3$ partition function of 
$\hat D$-quivers and linear quivers as the partition function of a gas of free 
fermions on a half line.

\subsection{Free fermions formalism for $\hat A$ quivers}
\label{sec:Afermi}

\begin{figure}[ht]
\centering
\epsfig{file=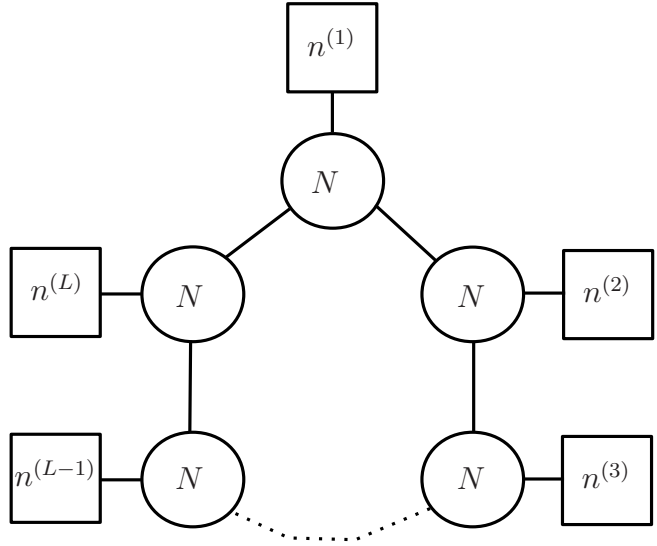,width=8.5cm
\psfrag{N}{$N$}
\psfrag{n1}{$n^{(1)}$}
\psfrag{n2}{$n^{(2)}$}
\psfrag{n3}{$n^{(3)}$}
\psfrag{nl1}{$\hspace{-0.21cm}n^{(L-1)}$}
\psfrag{nl}{$n^{(L)}$}
}
\caption{\label{fig:A}%
An $\hat{A}_{L-1}$-quiver. $n^{(a)}$ denotes the number of fundamental hypermultiplets 
at the corresponding $U(N)$ node.}
\end{figure}

We start by reviewing how the free fermion formalism arises for $\cN=3$ $\hat A$ quiver 
theories with $U(N)$ gauge groups \cite{Marino2012}. 
The $\hat{A}_{L-1}$-quiver diagram is presented in figure~\ref{fig:A}.
The matrix model for these theories is given generically by 
\beq
Z(N) = \frac{1}{N!^L} \int \prod_{a=1}^L d^N \lambda^{(a)} 
\prod_i F^{(a)}(\lambda^{(a)}_i) 
\frac{\prod_{i<j} \sh^2 (\lambda^{(a)}_i- \lambda^{(a)}_j) }
{\prod_{i,j} \ch (\lambda^{(a)}_i - \lambda^{(a+1)}_j + m^{(a)}) }\,,
\eeq
where $\lambda^{(a)}_i$, $i=1,\cdots,N$ are the eigenvalues of the $a$\textsuperscript{th} node 
(with the identification $\lambda^{(L+1)} \equiv \lambda^{(1)}$) and $m^{(a)}$ are the 
bifundamental masses. The factor $\prod_i F^{(a)}(\lambda^{(a)}_i) $ represents all 
terms which depend only on single eigenvalues and includes contributions from the CS term with level 
$k^{(a)}$, the FI term with parameter $\zeta^{(a)}$ and $n^{(a)}$ fundamental 
hypermultiplets with individual masses $\mu^{(a) }_\alpha$, $\alpha = 1,\cdots,n^{(a)} $ 
\beq
\label{FdefUN}
F^{(a)}(\lambda) 
= e^{\pi i k^{(a)}
\lambda^{2}} \,
e^{2\pi i \zeta^{(a)}
\lambda} 
\prod_{\alpha =1}^{n^{(a)}} \frac{1}{\ch(\lambda + \mu^{(a) }_\alpha )}\,.
\eeq
Using the Cauchy determinant identity
\beq
\label{cauchy}
\frac{\prod_{i<j} \sh(\lambda_i - \lambda_j) \sh(\tilde\lambda_i - \tilde\lambda_j)}
{\prod_{i, j}\ch(\lambda_i - \tilde\lambda_j)}
= \sum_{\sigma \in S_N}(-1)^{\sigma}\prod_{i=1}^N 
\frac{1}{\ch(\lambda_i - \tilde\lambda_{\sigma(i)})}\,,
\eeq
the partition function can be re-expressed as a sum over $L$ permutations
\beq
Z(N)= \frac{1}{N!^L} \sum_{\sigma^{(a)} \in S_N} (-1)^{\sum_{a=1}^L \sigma^{(a)}} 
\int\prod_{a=1}^L d^N \lambda^{(a)} 
\prod_i\frac{ F^{(a)}(\lambda^{(a)}_i ) }
{\ch (\lambda^{(a)}_i - \lambda^{(a+1)}_{\sigma^{(a)}(i)} + m^{(a)} )}\,.
\eeq
By relabelling the eigenvalues one can factor out all but one
of the permutations, picking up an overall factor of $N!^{L-1}$. This gives
\bal
\label{ZA}
Z(N)&= \frac{1}{N!} \sum_{\sigma \in S_N} (-1)^{\sigma} \int\prod_{a=1}^L d^N \lambda^{(a)} 
\prod_i^N 
\left( \prod_{a=1}^{L-1} 
\frac{ F^{(a)}(\lambda^{(a)}_i ) }{\ch (\lambda^{(a)}_i - \lambda^{(a+1)}_{i} + m^{(a)})}\right) 
\\&\hskip2.5in{}\times
\frac{ F^{(L)}(\lambda^{(L)}_i ) }{\ch (\lambda^{(L)}_i - \lambda^{(1)}_{\sigma(i)}+ m^{(L)} )}\,.
\eal
This integrand is a series of kernels of pairs of specific eigenvalues of successive nodes 
ultimately coupling each $\lambda^{(1)}_i$ with $\lambda^{(1)}_{\sigma(i)}$. 
This can be encoded graphically by the following diagram
\beq
\label{Aflow}
\left\{\lambda^{(1)} \right\} \rightarrow \left\{\lambda^{(2)} \right\} \rightarrow \cdots 
\rightarrow \left\{\lambda^{(L)} \right\} \overset{\sigma}{\rightarrow } \left\{\lambda^{(1)} \right\}\,.
\eeq

One can express the kernels in terms of canonical position and
momentum operators $\hat q$, $\hat p$, which satisfy 
$[\hat q, \hat p ] = i \hbar $. Taking $\lambda$ to be position eigenvalues, 
we have
\beq
\label{opident}
F(\lambda) \delta(\lambda'- \lambda) = \bra{\lambda'} F(\hat q) \ket{\lambda}\,, \quad 
\frac{1}{\ch (\lambda - \lambda')}=\bra{\lambda } \frac{1}{\ch \hat p} \ket{\lambda'}\,, \quad 
e^{2 \pi i m \hat p} \ket{\lambda} = \ket{\lambda - m}\,. 
\eeq
Here we used the standard relation between the position and momentum bases 
$\ket{p} = \int \frac{d\lambda}{\sqrt{2\pi \hbar}} 
\, e^{\frac{i p \lambda}{\hbar}} \ket{\lambda}$ and we have 
$\bra{\lambda_1} \lambda_2 \rangle = \delta(\lambda_1- \lambda_2)$ 
and $\bra{p_1} p_2 \rangle = \delta(p_1-p_2)$.
We choose to normalize $\hat p$ such that $\hbar = \frac{1}{2 \pi}$.
This allows one to write the integrand of \eqn{ZA} as
\beq
\bra{\lambda^{(1)}_i}F^{(1)}(\hat q )\frac{e^{2\pi im^{(1)}\hat p}}{\ch \hat p}\ket{\lambda^{(2)}_i}
\bra{\lambda^{(2)}_i}F^{(2)}(\hat q )\frac{e^{2\pi im^{(2)}\hat p}}{\ch \hat p}\ket{\lambda^{(3)}_i}
\cdots
\bra{\lambda^{(L)}_i}F^{(L)}(\hat q )\frac{e^{2\pi im^{(L)}\hat p}}{\ch \hat p}\ket{\lambda^{(1)}_{\sigma(i)}}\,.
\eeq
We obtain the final expression for $Z(N)$,
\beq
\label{ZAfermi}
Z(N)= \frac{1}{N!} \sum_{\sigma \in S_N} (-1)^{\sigma} \int d^N \lambda 
\prod_{i=1}^N \bra{\lambda_i} \hat\rho \ket{\lambda_{\sigma(i)}}
\eeq
with
\beq
\hat\rho = F^{(1)}(\hat q) \frac{e^{2 \pi i m^{(1)}\hat p} }{\ch\hat p}
F^{(2)}(\hat q) \frac{e^{2 \pi i m^{(2)}\hat p} }{\ch\hat p}
\cdots
F^{(L)}(\hat q) \frac{e^{2 \pi i m^{(L)}\hat p} }{\ch\hat p} \, .
\label{rhoAgen}
\eeq
This expression coincides with the partition function of $N$ non-interacting fermions%
\footnote{The fermionic statistics is understood from the anti-symmetrization over 
permutations of the positions $\lambda_i$.} 
living on a line, with a Hamiltonian $\hat H$ given by $\hat\rho = e^{-\hat H}$. 
In this language, $\lambda_i$ is the position of the $i$\textsuperscript{th} fermion on the line.

Clearly \eqn{ZAfermi} is fully determined by the spectrum of $\hat\rho$.
In this formulation there is thus 
a natural splitting of the computation of the partition 
function into two distinct steps. The first is to find a suitable density 
operator that encodes the desired quiver theory, and the second is to 
solve the resulting quantum mechanics problem. In this and the next section we 
concern ourselves only with finding the density operators and 
studying relations between them, we deal with the explicit 
computation of the partition function in section~\ref{sec:partition} 

For the remainder of this section we suppress the hats above operators.

\subsection{ $\hat D$-quiver analysis}

\begin{figure}[ht]
\centering
\epsfig{file=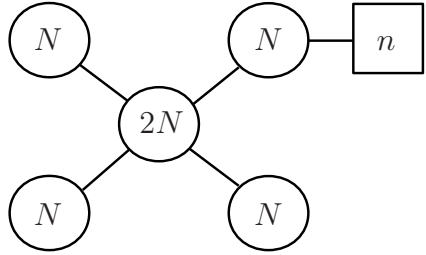,width=5.5cm
\psfrag{2N}{$2N$}
\psfrag{N}{$N$}
\psfrag{Nf}{$\,\, n$}
}
\caption{\label{fig:D4}%
A $\hat{D}_4$-quiver with $n$ fundamental hypermultiplets on one of the $U(N)$ nodes}
\end{figure}

We now implement
the Fermi-gas formulation to the matrix models computing the $S^3$
partition functions of $\hat D_{L+2}$-quiver theories with unitary gauge groups. 
These theories
are characterised by gauge group $U(2N)^{L-1} \times U(N)^4$ 
and a quiver structure as in figure~\ref{fig:Dgen}. 
We work out the full details in a simple example,
a $\hat D_4$-quiver theory with $n$ fundamental hypermultiplets attached to a single 
$U(N)$ node. We do not turn on mass or FI deformations. The quiver diagram for 
this theory is shown in figure~\ref{fig:D4}. The generalization to other 
$\hat D_L$-quivers is outlined in section~\ref{sec:longer}.

Our aim is now to find a suitable density operator that repackages the matrix model 
for this theory into an expression like \eqn{ZAfermi}. In order to do this it is
useful to collect the eigenvalues of pairs of terminal $U(N)$ into a single 
set of $2N$ eigenvalues. Anticipating this, we label the eigenvalues 
of one pair of $U(N)$ nodes respectively by%
\footnote{\label{notation abuse}This is a slight abuse of our earlier notation
where superscripts distinguished gauge group factors.} 
$\lambda^{(0)}_i$ and
$\lambda^{(0)}_{N+i}$, $ i = 1,\cdots, N$, and the other pair by
$\lambda^{(2)}_i$, $\lambda^{(2)}_{N+i}$.
The eigenvalues of the 
$U(2N)$ node are labelled \text{$\lambda^{(1)}_I$, $I = 1, \cdots, 2N$}. Following the rules
from appendix~\ref{sec:matrixmodels}, the matrix model for this theory is given by 
\bal
\label{ZD4}
Z(N)& = \frac{1}{N!^4 (2N)! } \int \prod_{a=0}^2 d^{2N} \lambda^{(a)} \prod_{I<J} 
\sh^2 \big(\lambda^{(1)}_I - \lambda^{(1)}_J \big) 
\\&
\frac{ \prod_{i<j} 
\sh^2 \big(\lambda^{(0)}_{i} -\lambda^{(0)}_{j} \big)
\sh^2 \big(\lambda^{(0)}_{N+i} -\lambda^{(0)}_{N+j} \big) 
\sh^2 \big(\lambda^{(2)}_{i} -\lambda^{(2)}_{j} \big)
\sh^2 \big(\lambda^{(2)}_{N+i} -\lambda^{(2)}_{N+j} \big) }
{\prod_{i,J} \ch \big(\lambda^{(0)}_{i} -\lambda^{(1)}_{J} \big)
\ch \big(\lambda^{(0)}_{N+i} -\lambda^{(1)}_{J} \big)\ch \big(\lambda^{(2)}_{i} -\lambda^{(1)}_{J} \big)
\ch \big(\lambda^{(2)}_{N+i} -\lambda^{(1)}_{J} \big) \prod_i \ch^{n}{\lambda^{(2)}_i }}\,.
\eal
Including artificially the factors $\prod_{i,j} 
\sh\big(\lambda^{(0)}_{i} -\lambda^{(0)}_{N+j} \big) \prod_{i,j} 
\sh\big(\lambda^{(2)}_{i} -\lambda^{(2)}_{N+j} \big)$ in the numerator and denominator, 
one can use the Cauchy identity \eqref{cauchy}, as well as a modified version
\beq
\label{cauchySh}
\frac{\prod_{i<j} \sh(\lambda_i - \lambda_j) \sh(\tilde\lambda_i - \tilde\lambda_j)}
{\prod_{i, j}\sh(\lambda_i - \tilde\lambda_j)}
= (-1)^{\frac{N(N-1)}{2}}\sum_{\sigma \in S_N}(-1)^{\sigma}\prod_{i=1}^N 
\frac{1}{\sh(\lambda_i - \tilde\lambda_{\sigma(i)})}\,,
\eeq
to re-expressed $Z(N)$ as
\bal
\label{ZD4cauchya}
Z (N) &= \frac{1}{N!^4 (2 N)! } 
\sum_{\substack{\sigma^{(a)}\in S_N\\ \ \tau^{(a)}\in S_{2N}}}
(-1)^{\sigma^{(0)}+\sigma^{(2)} + \tau^{(0)}+\tau^{(2)}} \int \prod_{a=0}^{2} d^{2N} \lambda^{(a)} 
\frac{1}{\prod_i \ch^{n} \lambda^{(2)}_i } 
\\&\hskip-.6cm
\prod_{i=1}^N \frac{1 }{\sh\big(\lambda^{(0)}_{i} - \lambda^{(0)}_{N+\sigma^{(0)}(i)} \big)}
\frac{1 }{\sh\big(\lambda^{(2)}_{i} - \lambda^{(2)}_{N+\sigma^{(2)}(i)} \big) } 
\prod_{I=1}^{2N} \frac{1}{\ch\big(\lambda^{(0)}_I - \lambda^{(1)}_{\tau^{(0)}(I) } \big) }
\frac{1}{\ch\big(\lambda^{(1)}_I - \lambda^{(2)}_{\tau^{(2)}(I) } \big) }\,.
\eal
Successive relabelings of the indices allow us to remove the sum 
over $\sigma^{(0)}$, $\sigma^{(2)}$ and $\tau^{(0)}$ and compensate for it by 
an overall factor of $N!^2 (2N)!$
\bal
\label{ZD4cauchy}
Z (N) &= \frac{1}{N!^2 } \sum_{\tau \in S_{2N} } (-1)^{ \tau} \int 
\prod_{a=0}^{2} d^{2N} \lambda^{(a)} 
\prod_{i=1}^N \frac{1}{\ch^{n} \lambda^{(2)}_i } 
\\&\quad \prod_{i=1}^N \frac{1 }{\sh\big(\lambda^{(0)}_{i} - \lambda^{(0)}_{N+i} \big) }
\frac{1 }{\sh\big(\lambda^{(2)}_{i} - \lambda^{(2)}_{N+i} \big) } \prod_{I=1}^{2N} 
\frac{1}{\ch\big(\lambda^{(0)}_I - \lambda^{(1)}_{I } \big) }
\frac{1}{\ch\big(\lambda^{(1)}_I - \lambda^{(2)}_{\tau(I) } \big) }\,.
\eal

As in the case of the $\hat A$-quivers \eqn{rhoAgen}, we would like to write this as the 
successive interaction between pairs of eigenvalues. Defining the reflection permutation 
$R$ by
\beq
R(i)=N +i\,, \qquad R(N+i)=i\,,
\eeq 
the integrand of the matrix model can be viewed as a series of kernels pairing eigenvalues 
of adjacent nodes in a chain that goes back and forth along the quiver, according to the 
diagram ({\em c.f.}, \eqn{Aflow})
\beq
\label{Dflow}
{\scriptstyle{R} \, }\text{\reflectbox{$\righttoleftarrow$}} \left\{ \lambda^{(0)}\right\} 
\rightleftarrows \left\{ \lambda^{(1)}\right\} \underset{\tau^{-1} }
{\overset{\tau}{\rightleftarrows }}\left\{ \lambda^{(2)}\right\} 
\text{\reflectbox{$\lefttorightarrow$}} { \scriptstyle{R}}\,.
\eeq

Traversing the quiver back and forth we end up with the composite permutation
\beq
R \tau^{-1} R \tau\,,
\eeq
so we can write the partition function in terms of a kernel relating $\lambda^{(1)}_I$ and 
$\lambda^{(1)}_{R\tau^{-1}R\tau(I)}$.
Note however that another eigenvalue of the central node $\lambda^{(1)}_{\tau^{-1}R\tau(I)}$ 
is integrated over to get this kernel. So for each permutation $\tau$ we need to choose half 
the eigenvalues of $\lambda^{(1)}$ on which the kernel acts. 
Let us call the set of $N$ indices of those eigenvalues $\cK(\tau)$. It is chosen 
to be closed under the composite permutation $R\tau^{-1}R\tau$ and such that 
$R$ takes this set to its complement $R(\cK(\tau))=\overline{\cK(\tau)}$. 
The partition function can be expressed in the following way
\bal
\label{ZD4K}
Z (N) &= \frac{1}{N!^2 } \sum_{\tau \in S_{2N} } (-1)^{ \tau} \int 
\prod_{a=0}^{2} d^{2N} \lambda^{(a)} 
\prod_{i=1}^N \frac{1}{\ch^{n} \lambda^{(2)}_i }
\prod_{k \in \cK(\tau)} \frac{1}{\ch\big(\lambda^{(1)}_k - \lambda^{(2)}_{\tau(k) } \big)} 
\\&\quad{}\times
\frac{(-1)^{s(\tau (k))} }{\sh\big(\lambda^{(2)}_{\tau(k)} - \lambda^{(2)}_{R \tau(k)} \big) }
\frac{1}{\ch\big(\lambda^{(2)}_{R \tau (k)} - \lambda^{(1)}_{\tau^{-1} R \tau (k)} \big)}
\frac{1}{\ch\big(\lambda^{(1)}_{\tau^{-1} R \tau (k)} - \lambda^{(0)}_{{\tau^{-1} R \tau (k)} } \big)}
\\&\quad{}\times
\frac{(-1)^{s(\tau^{-1} R\tau (k))} }
{\sh\big(\lambda^{(0)}_{\tau^{-1} R \tau (k)} - \lambda^{(0)}_{R\tau^{-1} R \tau (k)} \big) } 
\frac{1}{\ch\big(\lambda^{(0)}_{R\tau^{-1} R \tau (k)} - \lambda^{(1)}_{R\tau^{-1} R \tau (k) } \big) }\,,
\eal
where
\beq
s(k)=\begin{cases}
0\,,& k =1,\cdots, N\,,\\
1\,,&k = N+1,\cdots, 2N\,.
\end{cases}
\eeq

To be able to write the partition function in terms of a density operator we need
to include the contribution from the fundamental 
hypermultiplets into the product over $k$ in \eqn{ZD4K}. 
However, the fundamental hypermultiplets couple only to the eigenvalues 
$\lambda^{(2)}_i$ with $i=1,\cdots N$, which depending on $\tau$ is either 
$\lambda^{(2)}_{\tau(k)}$ {\it or} $\lambda^{(2)}_{R\tau(k)}$, but not both. 
These two options happen with equal probability for each combined 
permutation $R\tau^{-1}R\tau$, so we can write it as the sum (normalized by 
$1/2^N$)%
\footnote{In more detail, note that $R\tau^{-1}R\tau$ remains the same if one multiplies 
$\tau$ on the left by any combinations of two-cycles appearing in $R$. 
For a given $\tau$ this generates a set of $2^N$ terms, 
half with $\tau(k)$ in $\{1,\cdots,N\}$ and half in the compliment.}

\bal
\label{ZD4cauchyalltau}
Z (N) &= \frac{1}{2^N N!^2 } \sum_{\tau \in S_{2N} } (-1)^{ \tau} \int 
\prod_{a=0}^{2} d^{2N} \lambda^{(a)} 
\prod_{k \in \cK(\tau)} (-1)^{s (k) + s(\tau (k)) +1 }
\\&\quad{}
\prod_{k \in \cK(\tau)} \Bigg[\frac{1}{\ch\big(\lambda^{(1)}_k - \lambda^{(2)}_{\tau(k) } \big)} 
\frac{1}{\sh\big(\lambda^{(2)}_{\tau(k)} - \lambda^{(2)}_{R \tau(k)} \big) }
\frac{1}{\ch\big(\lambda^{(2)}_{R \tau (k)} - \lambda^{(1)}_{\tau^{-1} R \tau (k)} \big)}
\\&\hskip1.6cm{}
+ \frac{1}{\ch\big(\lambda^{(1)}_k - \lambda^{(2)}_{R\tau(k) } \big)} 
\frac{1}{\sh\big(\lambda^{(2)}_{R\tau(k)} - \lambda^{(2)}_{ \tau(k)} \big) }
\frac{1}{\ch\big(\lambda^{(2)}_{\tau (k)} - \lambda^{(1)}_{\tau^{-1} R \tau (k)} \big)}\Bigg] 
\frac{1}{\ch^{n} \lambda^{(2)}_{\tau (k)} } 
\\&\hskip-.7cm{}
\times
\frac{1}{\ch\big(\lambda^{(1)}_{\tau^{-1} R \tau (k)} - \lambda^{(0)}_{{\tau^{-1} R \tau (k)} } \big)}
\frac{1 }{\sh\big(\lambda^{(0)}_{\tau^{-1} R \tau (k)} - \lambda^{(0)}_{R\tau^{-1} R \tau (k)} \big) } 
\frac{1}{\ch\big(\lambda^{(0)}_{R\tau^{-1} R \tau (k)} - \lambda^{(1)}_{R\tau^{-1} R \tau (k) } \big) }\,,
\eal
where we have used 
$\prod_{k \in \cK(\tau)} (-1)^{s (\tau^{-1} R \tau (k))} = \prod_{k \in \cK(\tau)} (-1)^{s (k)+1}$.
This expression can now be recast as
\beq
\label{ZD4fermi}
Z(N)= \frac{1}{2^{2N} N!^2 } \sum_{\tau \in S_{2N} } (-1)^{ \tau} \int d^N \lambda 
\prod_{k \in \cK(\tau)} (-1)^{s (k) + s(\tau (k)) }
\rho(\lambda_k, \lambda_{R \tau^{-1} R \tau (k)})\,,
\eeq
with
\bal
\label{rhoD4pos}
\rho(\lambda, \lambda') = 2 \int \prod_{a=1}^5 d \lambda_a & 
\frac{-1}{\ch (\lambda - \lambda_1 )} \left(\frac{1}{\ch^{n} \lambda_1} 
\frac{1}{\sh (\lambda_1 -\lambda_2)} + \frac{1}{\sh (\lambda_1 -\lambda_2)} 
\frac{1}{\ch^{n} \lambda_2} \right) 
\\& \frac{1}{\ch (\lambda_2 - \lambda_3 )}\frac{1}{\ch (\lambda_3 - \lambda_4 )} 
\frac{1}{\sh (\lambda_4 - \lambda_5 )} \frac{1}{\ch (\lambda_5 - \lambda' )}\,,
\eal 
where we chose the normalization factor for convenience.

The kernel $\rho$ defines a density operator through the relation $\rho(\lambda_1,\lambda_2) = 
\bra{\lambda_1} \rho \ket{\lambda_2}$, which has a representation in terms of 
canonical position and momentum operators \eqn{opident}
\beq
\label{rhoD4}
\rho = \frac{1}{2} \frac{1}{\ch p} \left( \frac{1}{\ch^{n} q}
\frac{\sh p}{\ch p} + \frac{\sh p}{\ch p}\frac{1}{\ch^{n} q} \right)
\frac{\sh p}{ \ch^4 p}\,,
\eeq
where in addition to \eqn{opident} we have used 
\beq
\label{sinhfourier}
\frac{1 }{\sh (\lambda-\lambda')} 
=-\frac{i}{2} \bra{\lambda}\frac{\sh p}{\ch p } \ket{\lambda'}\,.
\eeq

To make further progress, we need to study the combinatorics of the composite 
permutations $R\tau^{-1} R \tau$. We relegate these additional technical calculations 
to appendix~\ref{sec:RRtauresults} and provide the final simplified result, which 
involves only a sum over permutations of $S_N$
\beq
Z(N) = \frac{1}{N!}\sum_{\sigma \in S_N} \frac{(-1)^\sigma}{2^{n_\sigma}}
\int \prod_{i=1}^N d\lambda_i \,\prod_{i=1}^N \rho(\lambda_i, \lambda_{\sigma(i)})\,,
\label{ZNgeneral}
\eeq
where $n_\sigma$ is the number of cycles in $\sigma$.

Because of the factor $1/2^{n_\sigma}$, \eqn{ZNgeneral} 
cannot be interpreted directly as the partition function of $N$ non-interacting fermions. 
When all FI and mass parameters are turned off, 
we find that it can be understood as resulting 
from a projection onto half of the states of 
a fermionic system. We show in appendix~\ref{sec:degenspectrum} that the density operator $\rho$ 
commutes with the reflection operator $\hat R$, defined by
\beq
\hat R \ket{\lambda} = \ket{-\lambda} \, .
\eeq
Consequently, the Hilbert space can be split into even and odd eigenstates. 
Furthermore, the spectra of even and odd eigenstates are identical, allowing us to 
rewrite $Z(N)$, using the projector $\frac{1 + \hat R}{2}$, as
\beq
\label{ZNprojplus}
Z(N)= \frac{1}{N!}\sum_{\sigma \in S_N} (-1)^\sigma 
\int \prod_{i=1}^N d\lambda_i \,\prod_{i=1}^N 
\bra{\lambda_i} \rho \bigg( \frac{1 + \hat R}{2} \bigg) \ket{\lambda_{\sigma(i)}}\,.
\eeq
This can be readily interpreted as the partition function of $N$ 
non-interacting fermions 
at positions $|\lambda_i|$ on a half-line with a Hamiltonian $H = -\log\rho$ 
where the operator $\frac{1 + \hat R}{2}$
is responsible for the projection onto particle states with even wavefunction on the line, or 
equivalently particle states on a half-line with Neumann boundary conditions.

Likewise we can use the projector $\frac{1 - \hat R}{2}$ 
to express the partition function in terms of the odd states
\beq
\label{ZNprojminus}
Z(N)= \frac{1}{N!}\sum_{\sigma \in S_N} (-1)^\sigma 
\int \prod_{i=1}^N d\lambda_i \,\prod_{i=1}^N 
\bra{\lambda_i} \rho \bigg( \frac{1 - \hat R}{2} \bigg) \ket{\lambda_{\sigma(i)}}\,.
\eeq
In this case we would interpret $Z(N)$ as the partition function of $N$ 
non-interacting fermions on a half-line with Dirichlet boundary condition at the origin. 

We did not find a free fermion interpretation of the partition function \eqref{ZNgeneral} 
for the cases with non-vanishing masses and FI parameters.


\subsection{Linear quiver analysis} 

We turn now to studying the matrix models of $U(2N)$ linear quivers terminating 
at each end with either an $Sp(2N)$ node or an antisymmetric hypermultiplet.
In many cases these quivers are known to be the mirror duals of 
$\hat D$-quivers \cite{Boer1997a}.

\begin{figure}[ht]
\centering
\epsfig{file=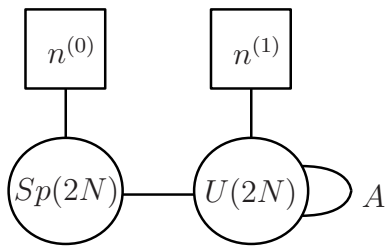, width=5cm
\psfrag{n0}{$n^{(0)}$}
\psfrag{n1}{$n^{(1)}$}
\psfrag{A}{$A$}
\psfrag{2N}{$\!\!\!U(2N)$}
\psfrag{Sp}{$\!\! Sp(2N)$}
}
\caption{A linear quiver with an $Sp(2N)$ node at one end 
and an antisymmetric hypermultiplet at the other.}
\end{figure}

We perform the analysis in the case of the quiver with 
a single $U(2N)$ node which is coupled to a single $Sp(2N)$ node and has also an
antisymmetric hypermultiplet. This example contains all the ingredients to treat 
any other quiver of this class, as we elaborate on in section~\ref{sec:longer}.

The $Sp(2N)$ node in our example has eigenvalues labelled 
$\lambda^{(0)}_i$, $i =1, \cdots, N$ and $n^{(0)}$
fundamental hypermultiplets. The $U(2N)$ node has eigenvalues 
$\lambda^{(1)}_I$, $I =1, \cdots, 2N$ an antisymmetric hypermultiplet 
as well as $n^{(1)}$ fundamental hypermultiplets. To simplify the expressions 
we do not include any FI terms or masses. 
Following the rules in appendix~\ref{sec:matrixmodels}, the matrix model is given by
\bal
\label{ZSpU}
Z(N) &= \frac{1}{2^{N} N! (2N)!} \int d^N \lambda^{(0)} d^{2N} \lambda^{(1)} 
\prod_{i=1}^N \frac{\sh^2 2 \lambda_i^{(0)}}
{\ch^{2 n^{(0)}} \lambda_i^{(0)}} \prod_{I=1}^{2N} \frac{1}{\ch^{n^{(1)}} \lambda^{(1)}_I}
\\&\qquad{}\times
\frac{ \prod_{i<j} \sh^2 \big(\lambda_i^{(0)}-\lambda_j^{(0)} \big) 
\sh^2 \big(\lambda_i^{(0)}+\lambda_j^{(0)} \big) 
\prod_{I<J} \sh^2 \big(\lambda_I^{(1)}-\lambda_J^{(1)}\big) }
{\prod_{i,J} \ch \big(\lambda^{(0)}_i -\lambda^{(1)}_J \big)
\ch \big(\lambda^{(0)}_i +\lambda^{(1)}_J \big) 
\prod_{I<J} \ch \big(\lambda^{(1)}_I +\lambda^{(1)}_J \big) }\,.
\eal

A first step is to write the contribution of the $Sp$ node in terms of $2N$ eigenvalues satisfying
\beq
\label{Splambda}
\lambda^{(0)}_{N+i} = - \lambda^{(0)}_i\,.
\eeq
The interaction between the $Sp(2N)$ and $U(2N)$ nodes 
combine to a single Cauchy determinant \eqref{cauchy}
\bal
\label{cauchySpU}
&\frac{\prod_{i<j} \sh^2 \big(\lambda^{(0)}_i - \lambda^{(0)}_j\big) 
\sh^2 \big(\lambda^{(0)}_i + \lambda^{(0)}_j\big) 
\prod_{i=1}^N \sh 2 \lambda^{(0)}_i \prod_{I<J} \sh \big(\lambda^{(1)}_I - \lambda^{(1)}_J \big)}
{\prod_{i,J} \ch \big(\lambda^{(0)}_i -\lambda^{(1)}_J \big)
\ch \big(\lambda^{(0)}_i +\lambda^{(1)}_J \big)} 
\\& \hspace{.5cm}
= \frac{\prod_{I<J} \sh \big(\lambda^{(0)}_I - \lambda^{(0)}_J\big)
\sh \big(\lambda^{(1)}_I - \lambda^{(1)}_J\big)}
{\prod_{I,J} \ch \big(\lambda^{(0)}_I -\lambda^{(1)}_J \big) }
= \sum_{\tau^{(0)} \in S_{2N} } (-1)^{\tau^{(0)}} \prod_{I=1}^{2N} \frac{1}
{\ch \big(\lambda^{(0)}_I - \lambda^{(1)}_{\tau^{(0)}(I)}\big) }\,.
\eal
The remaining terms involving the eigenvalues of the $U(2N)$ node can be 
interpreted as a Pfaffian, rather than a determinant. We can use the identity 
\cite{Dey2013, Kuperberg2002} 
\beq
\prod_{I<J \leq 2N} \frac{x_I - x_J}{1 + x_I x_J }=
\Pf\left(\frac{x_I - x_J }{1 + x_I x_J }\right) = 
\frac{1}{ 2^N N!} \sum_{ \tau \in S_{2N} } (-1)^\tau \prod_{i =1}^N 
\frac{x_{\tau(i)} - x_{\tau R(i)} }{1 + x_{\tau(i)} x_{\tau R(i)} }\,,
\eeq
where $R$ is again the permutation $R(i) = N+i$ modulo $2N$. 
Plugging $x = e^{2\pi \lambda^{(1)}}$, 
we obtain 
\beq
\label{pfaffSpU}
\prod_{I<J} \frac{\sh \big(\lambda^{(1)}_I - \lambda^{(1)}_J\big)}
{\ch \big(\lambda^{(1)}_I + \lambda^{(1)}_J\big)}=
\frac{1}{ 2^N N!}\sum_{\tau \in S_{2N}} (-1)^\tau\prod_{i=1}^N 
\frac{\sh \big(\lambda^{(1)}_{\tau(i)} - \lambda^{(1)}_{\tau R(i)} \big)}
{\ch \big(\lambda^{(1)}_{\tau(i)} + \lambda^{(1)}_{\tau R(i)} \big)}\,.
\eeq

As before, we can remove one of the permutations coming from
\eqn{cauchySpU} and \eqn{pfaffSpU} by a relabelling of eigenvalues, 
picking up an overall factor of $(2N)!$. This gives
\bal
Z(N) = \frac{1}{2^{2N} N!^2} &\sum_{\tau \in S_{2N}} (-1)^\tau
\int d^N \lambda^{(0)} d^{2N} \lambda^{(1)} 
\prod_{i=1}^N \frac{\sh 2 \lambda_i^{(0)}}
{\ch^{2 n^{(0)}} \lambda_i^{(0)}} \prod_{I=1}^{2N} \frac{1}{\ch^{n^{(1)}} \lambda^{(1)}_I}
\\&
\prod_{i=1}^N 
\frac{1}{\ch \big(\lambda^{(0)}_{\tau(i)} - \lambda^{(1)}_{\tau(i)}\big) } 
\frac{\sh \big(\lambda^{(1)}_{\tau(i)} - \lambda^{(1)}_{\tau R(i)} \big)}
{\ch \big(\lambda^{(1)}_{\tau(i)} + \lambda^{(1)}_{\tau R(i)} \big)}
\frac{1}{\ch \big(\lambda^{(1)}_{\tau R(i)}-\lambda^{(0)}_{\tau R(i)}\big) } 
\,.
\eal
replacing for convenience $\tau\to\tau^{-1}$, 
we can again rewrite the expression as a product over 
the set $\cK(\tau)$, consisting of $N$ indices 
closed under the permutation $R \tau^{-1} R \tau$
\bal
Z(N) &= \frac{1}{2^{2N} N!^2} \sum_{\tau \in S_{2N}} (-1)^\tau
\int d^N \lambda^{(0)} d^{2N} \lambda^{(1)} 
\prod_{k\in \cK(\tau)} \frac{(-1)^{s(k) } \sh 2 \lambda_k^{(0)}}{\ch^{2 n^{(0)}} \lambda_k^{(0)}} 
\frac{1}{\ch^{n^{(1)}} \lambda^{(1)}_k\ch^{n^{(1)}} \lambda^{(1)}_{R(k)}}
\\&\quad{}
\frac{1}{\ch \big(\lambda^{(0)}_{k} - \lambda^{(1)}_{k}\big) } 
\frac{(-1)^{s(\tau(k) ) } \sh \big(\lambda^{(1)}_{k} - \lambda^{(1)}_{\tau^{-1} R\tau(k)} \big)}
{\ch \big(\lambda^{(1)}_{k} + \lambda^{(1)}_{\tau^{-1} R\tau(k)} \big)}
\frac{1}{\ch \big(\lambda^{(1)}_{\tau^{-1}R\tau(k)}+\lambda^{(0)}_{R\tau^{-1}R\tau(k)}\big) } 
\,.
\eal
The $(-1)^{s(\tau(k) )}$ signs comes from re-expressing 
$\sh \big(\lambda^{(1)}_{\tau(i)} - \lambda^{(1)}_{\tau R(i)} \big)$
in terms of $k$, while
\eqn{Splambda} is responsible for the $(-1)^{s(k)}$ signs as well as allowing
the replacement in the last denominator
\beq
\lambda^{(0)}_{ \tau^{-1} R\tau(k)} 
=-\lambda^{(0)}_{ R\tau^{-1} R\tau(k)}\,.
\eeq
As in the case of the $\hat D$-quivers \eqref{ZD4fermi}, we obtain a density operator 
between two $\lambda^{(0)}$ eigenvalues related by the permutation $R\tau^{-1}R\tau$
\beq
\label{rhoSpUposs}
\rho(\lambda, \lambda') = \int d \lambda_1d\lambda_2\,
\frac{\sh{ 2 \lambda }}{\ch^{2n^{(0)}} \lambda} \frac{1}{\ch (\lambda - \lambda_1)}
\frac{1}{\ch^{n^{(1)}} \lambda_1} \frac{\sh(\lambda_1-\lambda_2)}{\ch(\lambda_1+\lambda_2)}
\frac{1}{\ch^{n^{(1)}} \lambda_2} \frac{1}{\ch(\lambda_2+\lambda')}\,,
\eeq
in terms of which the partition function is given exactly as in \eqn{ZD4fermi}. 
Expanding $\sh(\lambda_1-\lambda_2)$ and reversing the sign of $\lambda_2$ allows us again to 
represent the operator in terms of canonical position and momentum operators
\bal
\label{rhoSpUop}
\rho &= \frac{1}{2} \frac{\sh{ 2 q }}{\ch^{2n^{(0)}} q} \frac{1}{\ch p} \frac{1}{\ch^{n^{(1)}} q} 
\left( \sh q \frac{1}{\ch p} \ch q + \ch q \frac{1}{\ch p} \sh q \right) 
\frac{1}{\ch^{n^{(1)}} q} \frac{1} {\ch p}
\\&= \frac{1}{2} \frac{\sh{ 2 q }}{\ch^{2n^{(0)}} q} \frac{1}{\ch p} \frac{1}{\ch^{n^{(1)}} q} 
\left( e^{\pi q} \frac{1}{\ch p}e^{\pi q} +e^{-\pi q}\frac{1}{\ch p}e^{-\pi q}\right) 
\frac{1}{\ch^{n^{(1)}} q} \frac{1} {\ch p}\,.
\eal

The same arguments as for the $\hat D$-quiver allow us to express $Z(N)$ as the partition 
function of $N$ non-interacting fermions on a half line with Neumann \eqref{ZNprojplus} 
or Dirichlet \eqref{ZNprojminus} boundary conditions. 
Indeed, as we show in section~\ref{sec:mirror}, the density operators of 
many linear quivers are related to those of $\hat D$-quivers by a linear canonical transformation.

\subsection{Generalization to longer quivers}
\label{sec:longer}

It is straight-forward to generalize the analysis 
to quivers with an arbitrary number of $U(2N)$ nodes and arbitrary number of fundamental 
hypermultiplets on each node. Just as for the
$\hat A$-quiver theories, the matrix model contributions from hypermultiplets transforming in 
the bifundamental representation of pairs of $U(2N)$ gauge nodes combine with the 
vector multiplet contributions 
to form one Cauchy determinant between each pair of adjacent nodes. This translates into 
$\ch^{-1}p$ terms in the density operator. We represent the contribution from fundamental 
hypermultiplets and the FI and CS terms of the $a$\textsuperscript{th} node again by 
$F^{(a)}( q)$ \eqn{FdefUN}. This leads to a piece in the density operator of the form
\beq
\label{U2Nfactors1}
\frac{1}{\ch p} F^{(1)}( q) \frac{1}{\ch p} F^{(2)}( q) \frac{1}{\ch p} F^{(3)}( q) \frac{1}{\ch p}\cdots
\eeq
In all of our examples the density operator combines kernels 
going back and forth along the quiver. 
The contribution from the ends of the quivers ($U(N)$ nodes for 
$\hat D$-quivers and $Sp(2N)$/antisymmetric hypermultiplet for linear quivers) are 
the same as in the previous sections.

For the $\hat{D}$-quivers the contribution from going back along the quiver as in \eqref{Dflow} gives, 
\beq
\label{U2Nfactors2}
\cdots \frac{1}{\ch p} F^{(3)}( q) \frac{1}{\ch p} F^{(2)}( q) \frac{1}{\ch p} F^{(1)}( q) \frac{1}{\ch p}\,.
\eeq
For the linear quivers, the antisymmetric hypermultiplet or $Sp$ node introduce 
a minus sign, like the replacement $\lambda_2 \rightarrow - \lambda_2$
that gave \eqn{rhoSpUop} from \eqn{rhoSpUposs}. Therefore the second part of the density 
operator includes
\beq
\label{U2Nfactors3}
\cdots \frac{1}{\ch p} F^{(3)}( -q) \frac{1}{\ch p} F^{(2)}( -q) \frac{1}{\ch p} F^{(1)}(- q) \frac{1}{\ch p}\,.
\eeq

\subsubsection{Generalized $\hat D$-quivers}

Let us consider a $\hat{D}_{L+2}$ quiver with arbitrary number of gauge nodes and 
fundamental hypermultiplets on each node,%
\footnote{\label{badDquiv}Note that in order for the matrix model to be convergent, such a theory 
must have at least one fundamental hypermultiplet. 
With this condition violated the formal manipulations 
still go through, but the divergence of the matrix model 
translates to a density operator which is not of trace class.}
as shown in figure~\ref{fig:Dgen}.
We label the $U(2N)$ nodes by $1, \cdots, L-1 $, 
and as in figure~\ref{fig:Dgen}, we distinguish parameters from pairs of terminal 
$U(N)$ nodes by primes: $F^{(0)}$, $F^{\prime(0)}$,$F^{(L)}$, $F^{\prime(L)}$. 
We note that all bifundamental 
hypermultiplet masses can be set to zero by shifting eigenvalues.%
\footnote{\label{phase}Shifting eigenvalues to remove masses can 
introduce an overall phase in the matrix model. Such phases are unphysical, 
in the sense that they arise from background (mixed) Chern-Simons terms that can be added to 
the regularization scheme when computing the partition function of the 3d theories 
\cite{Closset:2012vg, Closset:2012vp}.} 
The above rules lead to the density operator
\bal
\label{rhoDgen}
\rho = &\frac{1}{4} \frac{1}{\ch p} \left( F^{(0)}( q) 
\frac{\sh p }{\ch p} F^{\prime(0)}( q) +F^{\prime(0)}( q) 
\frac{\sh p }{\ch p} F^{ (0)}( q) \right) \frac{1}{\ch p } \left( \prod_{a=1}^{L-1} 
F^{(a)}( q) \frac{1}{\ch p}\right)
\\&\hspace{1cm} \left( F^{(L)}( q) 
\frac{\sh p }{\ch p} F^{\prime(L)}( q) + F^{\prime(L)}( q) 
\frac{\sh p }{\ch p} F^{ (L)}( q) \right) \frac{1}{\ch p}\left(\prod_{a=1}^{L-1} 
F^{(L-a)}( q) \frac{1}{\ch p}\right). 
\eal
We can easily recover \eqn{rhoD4} by setting $L=2$, $F^{(2)}(q) = \ch^{-n} q$ 
and all other $F^{(a)}=F^{\prime(a)}=1$.

\subsubsection{Generalized linear quivers}

We can proceed in a similar fashion to write down the density operators for longer linear 
quivers, where each end of the $U(2N)$ linear chain has either an $Sp(2N)$ node
or an antisymmetric hypermultiplet and any number of fundamental hypermultiplets 
on all the nodes.%
\footnote{While the formal manipulations again go through in all cases 
(see also footnote~\ref{badDquiv}), convergence of the matrix model requires 
a total of at least three fundamental hypermultipets, with at least one coupling to 
the terminating nodes at each end of the quiver.}
Again we note that the masses for all bifundamental 
hypermultiplets between $U(2N)$ nodes 
can be set to zero by shifts of the eigenvalues. We cannot always do the same 
for the mass of antisymmetric hypermultiplets, or for those charged under 
$Sp(2N)$, so we keep these masses as well as the fundamental hypermultiplet masses.
We consider a quiver with $L$ $U(2N)$ nodes
and again package the FI, CS and fundamental hypermultiplet contributions into $F^{(a)}$.
Two instances (out of three) of such general linear quivers are pictured in figure~\ref{fig:Spgenfund}. 
The density operator is given by
\beq
\label{rhoSpAgen}
\rho = B^{(0)}( p, q) \left( \prod_{a=1}^{L-1} 
F^{(a)}( q) \frac{1}{\ch p}\right) F^{(L)}( q) B^{(L+1)}( p, q) 
\left( \prod_{a=1}^{L-1} F^{(L+1-a)}(- q) \frac{1}{\ch p}\right) F^{(1)}(- q)\,,
\eeq
where the functions $B^{(a)}(p, q)$ account for whether the ends of the quiver terminate 
with an $Sp(2N)$ node or antisymmetric hypermultiplet.

If the quiver terminates with an $Sp(2N)$ node we have
\beq
B^{(a)}_{Sp}(p, q) = \frac{e^{2 \pi i m^{(a)} p}}
{\ch p} \sh (2 q) \widetilde F^{(a)}( q) \frac{e^{2 \pi i m^{(a)} p}}{\ch p}\,, 
\eeq
where $m^{(a)}$ is the bifundamental mass and $\widetilde F^{(a)}(q)$ 
captures the contributions from the CS term with level $k^{(a)}$ and $n^{(a)}$ fundamental 
hypermultiplets of $Sp(2N)$ with masses $\mu^{(a)}_\alpha$
\beq
\widetilde F^{(a)}(q)
= \frac{e^{2 \pi i k^{(a)} q^2} }
{\prod_{\alpha=1}^{n^{(a)}} \ch(q - \mu^{(a)}_\alpha) \ch(q + \mu^{(a)}_\alpha)}\,.
\eeq

If it terminates with an antisymmetric hypermultiplet we have
\beq
B^{(a)}_{A}( p, q) = \frac{1}{2} \left( e^{\pi q} \frac{e^{2 \pi i M^{(a)} p}}{\ch p } e^{\pi q}
+ e^{-\pi q} \frac{e^{2 \pi i M^{(a)} p}}{\ch p } e^{-\pi q} \right),
\eeq
where $M^{(a)}$ is the antisymmetric mass.

Note that the expression \eqn{rhoSpAgen} assumes there is at least one $U(2N)$ node. 
There are two relevant cases without $U(2N)$ nodes:
For the single node $Sp(2N)$ theory with an antisymmetric hypermultiplet the density 
operator is given by%
\footnote{This theory was previously studied from the Fermi gas perspective in \cite{Mezei2014},
where a rather different density operator was obtained. We compare the different 
formalisms in appendix~\ref{sec:prevfermi}.} 
\beq
\label{rho1sp}
\rho = \frac{1}{2} \sh (2 q) \widetilde F^{(0)}( q)\left( \sh q \frac{e^{2 \pi i M p}}{\ch p } \ch q 
+ \ch q \frac{e^{2 \pi i M p}}{\ch p } \sh q \right).
\eeq
For the $Sp(2N) \times Sp(2N)$ theory the density 
operator is given by
\beq
\label{rho2sp}
\rho = \frac{e^{2 \pi i m^{(0)} p}}{\ch p} \sh (2 q) \widetilde F^{(0)}( q) 
\frac{e^{2 \pi i m^{(0)} p}}{\ch p} \sh (2 q) \widetilde F^{(1)}( q)\,.
\eeq

\section{Mirror symmetry}
\label{sec:mirror}

In \cite{Drukker2015} it was found that pairs of mirror dual $\hat A$-quiver theories give rise 
to Fermi gas formulations which are related to each other by simple 
linear canonical transformations acting on their density operators. In this way, once one 
understands how to go from a quiver theory to the Fermi gas density operator, one can with 
little effort find the mirror map between dual theories. In this section we show 
that this continues to hold true for 
the mirror pairs involving $\hat D$-quivers and linear quivers. 
In this section we consider theories with 
mass and FI term deformations, since those get mapped to each other 
under mirror symmetry. For simplicity we restrict our attention in the remainder of the 
paper to theories without Chern-Simons terms

We proceed with a number of examples, starting in each case with a $\hat D$-quiver, 
and demonstrating how the canonical transformation
\bal
\label{mirrortrans}
p \rightarrow q, \qquad q \rightarrow - p\,,
\eal 
maps the density operator onto that of the mirror linear quiver theory.

\subsection{$\hat{D}_4$-quiver with two fundamentals}

The first example we consider is the $\hat D_4$-quiver with a fundamental hypermultiplet 
on two of the terminal $U(N)$ nodes. This example is somewhat special as there are three
inequivalent ways of pairing up the terminal $U(N)$ nodes, which leads to 
different density operators. The canonical transformations of these three descriptions are 
related to three different mirror theories. 
The existence of several mirror dual theories was already noted in 
section~4.4.3 of \cite{Dey2014a}.
\begin{figure}[ht]
\centering
\epsfig{file=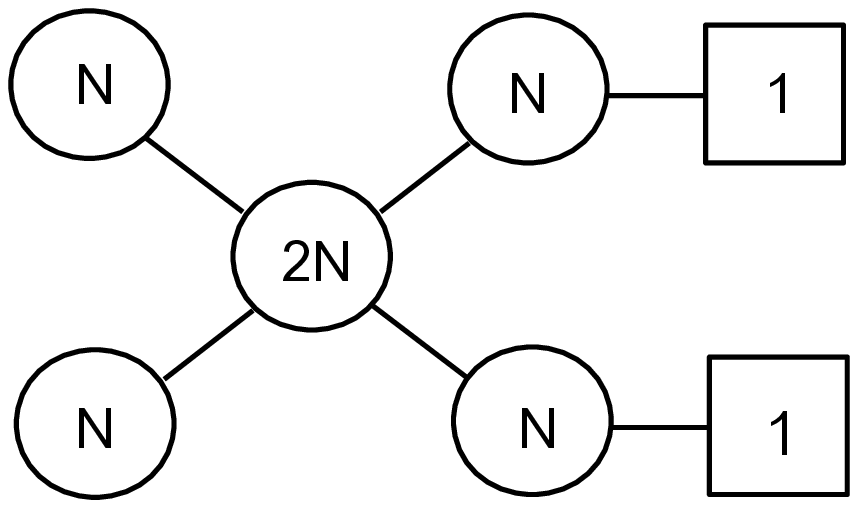,width=6cm
\psfrag{2}{$\!2$}
\psfrag{N}{$\!\!N$}
\psfrag{1}{$\!1$}
}
\hspace{1cm}
\epsfig{file=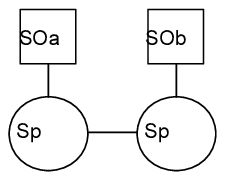,width=4.2cm
\psfrag{Sp}{$\! Sp(2N)$}
\psfrag{SOa}{$\hspace{0.5cm} 3$}
\psfrag{SOb}{$\hspace{0.5cm} 1$}
}
\caption{$\hat D_4$-quiver and its mirror dual theory with $Sp(2N) \times Sp(2N)$ gauge group. }
\label{D4pairing1}
\end{figure}

The first possibility is to pair the two terminal nodes without fundamental matter and the 
two with fundamental matter, as shown in figure~\ref{D4pairing1}. Within each pairing 
we distinguish the FI parameters of the two $U(N)$ nodes by giving one of them a prime. 
We do not turn on masses for the two $U(N)$ fundamental hypermultiplets.%
\footnote{One of these mass parameters can be removed by shifts of the matrix model
eigenvalues. The other is mapped under mirror symmetry to a
``hidden'' FI parameter \cite{Kapustin1998, Feng2000}, which does not have a clear 
interpretation in the mirror gauge theory.} 
The density operator of the $\hat D_4$ theory can be read from \eqn{rhoDgen}
\bal
\rho &= \frac{1}{4}\frac{1}{\ch p} \left( e^{2 \pi i \zeta^{(0)} q} 
\frac{ \sh p} {\ch p} e^{2 \pi i \zeta^{\prime(0)} q} +e^{2 \pi i \zeta^{\prime(0)} q} 
\frac{ \sh p} {\ch p} e^{2 \pi i \zeta^{(0)} q} \right)
\frac{1}{\ch p } e^{2 \pi i \zeta^{(1)} q} 
\\&\quad{}
\times\frac{1}{\ch p }\frac{1}{\ch q} \left(e^{2 \pi i \zeta^{(2)} q} 
\frac{ \sh p} {\ch p} e^{2 \pi i \zeta^{\prime(2)} q}
+e^{2 \pi i \zeta^{\prime(2)} q}\frac{ \sh p} {\ch p} e^{2 \pi i \zeta^{(2)} q} \right)
\frac{1}{\ch q}\frac{1}{\ch p} e^{2 \pi i \zeta^{(1)} q}\,.
\eal
To map this to the density operator of the mirror dual theory, we first use the relation
\beq
\label{shiftid}
e^{2\pi i \zeta q} f(p) e^{-2 \pi i \zeta q} = f(p- \zeta)\,,
\eeq
to simplify the terms in parenthesis
\bal
\label{mirrorid1} 
& e^{\pi i (\zeta-\zeta') q} \frac{ \sh p} {\ch p} e^{-\pi i(\zeta- \zeta') q} 
+e^{-\pi i (\zeta-\zeta') q} \frac{ \sh p} {\ch p} e^{\pi i( \zeta-\zeta') q} 
\\&=
\frac{ \sh (p+ \frac{1}{2} \zeta' - \frac{1}{2}\zeta) } 
{\ch ( p + \frac{1}{2} \zeta' - \frac{1}{2}\zeta )} + 
\frac{ \sh ( p-\frac{1}{2} \zeta'+ \frac{1}{2}\zeta) } 
{\ch (p -\frac{1}{2} \zeta'+ \frac{1}{2} \zeta)}
=\frac{ 2\sh 2 p}{\ch (p+ \frac{1}{2} \zeta - \frac{1}{2} \zeta' )
\ch (p-\frac{1}{2} \zeta + \frac{1}{2} \zeta' )}\,.
\eal
By further commuting exponential factors we get
\bal
\rho &= e^{\pi i(\zeta^{(0)}+\zeta^{\prime(0)})q}
\frac{1}{\ch \big(p+\tilde\mu^{(1)}_1\big)} 
\frac{\sh 2 p}{\ch\big(p+\tilde\mu^{(1)}_2\big)\ch\big(p-\tilde\mu^{(1)}_2\big)}
\frac{1}{\ch\big(p-\tilde\mu^{(1)}_1\big)} 
\frac{1}{\ch\big(p+\tilde\mu^{(1)}_3\big)}
\\&\quad{}\times
\frac{e^{-\pi i\widetilde m q}}{\ch q} 
\frac{\sh 2 p}{\ch\big(p+\tilde\mu^{(0)}\big)\ch\big(p-\tilde\mu^{(0)}\big)}
\frac{e^{-\pi i\widetilde m q}}{\ch q} 
\frac{1}{\ch\big(p-\tilde\mu^{(1)}_3\big)} 
e^{-\pi i(\zeta^{(0)}+\zeta^{\prime(0)})q}\,.
\eal
with
\bal
\quad \tilde \mu_1^{(1)} &= \tfrac{1}{2} (\zeta^{(0)}+\zeta^{\prime(0)})\,, 
\quad \tilde\mu_2^{(1)} = \tfrac{1}{2} (\zeta^{(0)}-\zeta^{\prime(0)})\,,
\quad \tilde\mu_3^{(1)} = -\zeta^{(1)}-\tfrac{1}{2} (\zeta^{(0)} + \zeta^{\prime(0)})\,, \\
\tilde\mu_{\phantom{1}}^{(0)} &= \tfrac{1}{2} (\zeta^{(2)} - \zeta^{\prime(2)})\,, \\
\widetilde{m} &= - \zeta^{(1)} - \tfrac{1}{2} ( \zeta^{(0)} +\zeta^{\prime(0)}+ \zeta^{(2)}+ \zeta^{\prime(2)} )\,.
\label{ParamMap1}
\eal
Now we can act on the density operator by canonical transformation \eqn{mirrortrans}. 
In addition we conjugate the operator to remove the exponential factors at the beginning 
and end, which clearly does not alter the spectrum. This gives
\beq
\label{SpSpmirror}
\tilde\rho = \frac{e^{2 \pi i \widetilde m p} }{\ch p} 
\frac{\sh 2 q}{\ch \big(q +\tilde\mu_{\phantom{1}}^{(0)}\big) 
\ch \big(q -\tilde\mu_{\phantom{1}}^{(0)}\big)} 
\frac{e^{2 \pi i \widetilde m p} }{\ch p} \frac{\sh 2 q}
{\prod_{\alpha=1}^3 \ch \big(q + \tilde\mu_\alpha^{(1)} \big) \ch \big(q - \tilde\mu_\alpha^{(1)} \big)}\,.
\eeq
As advertized we recover the density operator for a linear quiver with two $Sp(2N)$ nodes. 
One with a single fundamental hypermultiplet and the other with three \eqref{rho2sp}.
The relations between the FI and mass deformation parameters of the mirror dual theories are 
expressed in \eqref{ParamMap1}, where $\widetilde{m}$ is the bifundamental hypermultiplet mass 
and $\tilde\mu^{(0)}, \tilde\mu_\alpha^{(1)}$ are the fundamental hypermultiplet masses
of the dual theory. This mirror map generalises slightly the one found already in 
\cite{Dey2013}, allowing for $\zeta^{(0)}\neq\zeta^{\prime(0)}$ and 
$\zeta^{(2)}\neq\zeta^{\prime(2)}$, which translate to additional mass deformations 
in the dual linear quiver theory.

\begin{figure}[ht]
\centering
\epsfig{file=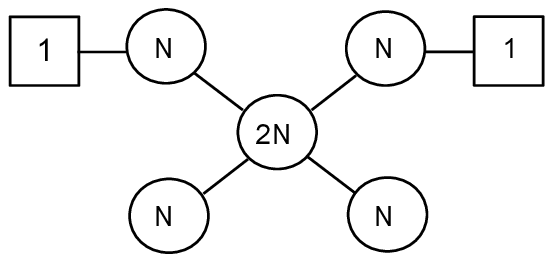,width=7.5cm
\psfrag{2N}{$2N$}
\psfrag{N}{$N$}
\psfrag{1}{$\!1$}
}
\hspace{1cm}
\epsfig{file=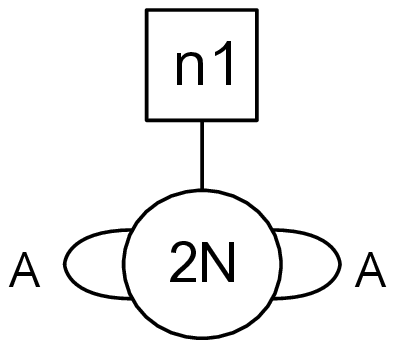,width=3.6cm
\psfrag{2}{\!\!\!\!$U(2N)$}
\psfrag{N}{}
\psfrag{A}{$A$}
\psfrag{n1}{$\ 4$}
}
\caption{$\hat D_4$-quiver and its mirror dual theory with $U(2N)$ gauge group and 
antisymmetric hypermultiplets. }
\label{D4pairing2}
\end{figure}
A second way of obtaining a density operator for the $\hat D_4$-quiver comes from pairing nodes 
with and without fundamental hypermultiplet, as in figure~\ref{D4pairing2}. After shifting 
eigenvalues to remove masses we are left with only one nonzero mass $\mu$ for one of the 
fundamental hypermultiplets. From \eqn{rhoDgen} we can write down the density operator 
\bal
\label{D4rhonoz2}
\rho &= \frac{1}{4} \frac{1}{\ch p} 
\left( \frac{e^{2 \pi i \zeta^{(0)}q } }{\ch q } 
\frac{\sh p}{\ch p} e^{2 \pi i \zeta^{\prime(0)}q } + e^{2 \pi i \zeta^{\prime(0)}q }
\frac{\sh p}{\ch p} \frac{e^{2 \pi i \zeta^{(0)} q } }{\ch q } \right) 
\frac{1}{\ch p} e^{2 \pi i \zeta^{(1)} q} 
\\&\quad{}\times 
\frac{1}{\ch p} \left( \frac{e^{2 \pi i \zeta^{(2)}q } }{\ch (q + \mu) } 
\frac{\sh p}{\ch p} e^{2 \pi i \zeta^{\prime(2)}q } + e^{2 \pi i \zeta^{\prime(2)}q }
\frac{\sh p}{\ch p} \frac{e^{2 \pi i \zeta^{(2)} q } }{\ch (q + \mu) } \right) 
\frac{1}{\ch p} e^{2 \pi i \zeta^{(1)} q}\,.
\eal

Once again we start by manipulating the expression of the density operator, 
using the relation \eqref{shiftid}
\begin{align}
\rho &= \frac{1}{4} e^{- i \pi \zeta^{(1)} q} \frac{1}{\ch( p + \frac{\zeta^{(1)}}{2} )} 
\frac{1}{\ch( p - \frac{\zeta^{(1)}}{2} )} 
\nonumber\\
& \times \left( \frac{e^{2 \pi i (\zeta^{(0)}+\zeta^{\prime(0)}+\zeta^{(1)}) q } }{\ch q } 
\frac{\sh (p+\zeta^{\prime(0)} + \frac{\zeta^{(1)}}{2})}{\ch(p+\zeta^{\prime(0)} + \frac{\zeta^{(1)}}{2})} 
+\frac{\sh (p-\zeta^{\prime(0)} - \frac{\zeta^{(1)}}{2})}{\ch (p-\zeta^{\prime(0)} - \frac{\zeta^{(1)}}{2})} 
\frac{e^{2 \pi i (\zeta^{(0)} +\zeta^{\prime(0)} + \zeta^{(1)}) q } }{\ch q } \right) 
\nonumber\\
& \times 
\frac{1}{\ch (p + \frac{\zeta^{(1)}}{2}) }
\frac{1}{\ch (p- \frac{\zeta^{(1)}}{2}) } 
\nonumber\\
& \times 
\left( \frac{e^{2 \pi i (\zeta^{(2)}+\zeta^{\prime(2)}+\zeta^{(1)}) q } }{\ch (q + \mu) } 
\frac{\sh (p+\zeta^{\prime(2)}+ \frac{\zeta^{(1)}}{2})}{\ch (p+\zeta^{\prime(2)}+ \frac{\zeta^{(1)}}{2})} 
+ \frac{\sh (p-\zeta^{\prime(2)}-\frac{\zeta^{(1)}}{2})}{\ch (p-\zeta^{\prime(2)}-\frac{\zeta^{(1)}}{2})} 
\frac{e^{2 \pi i (\zeta^{(2)} +\zeta^{\prime(2)}+\zeta^{(1)} ) q } }{\ch (q + \mu) } \right) 
\nonumber\\
& \times e^{ i \pi \zeta^{(1)} q}.
\end{align}
We now use the identity
\bal
\label{mirrorid2}
&\hskip-2cm\frac{e^{2 \pi i \zeta q}}{\ch (q+ \mu) }\frac{\sh p+\zeta'}{\ch p+\zeta'} + 
\frac{\sh p-\zeta'}{\ch p-\zeta'} 
\frac{e^{2 \pi i \zeta q}}{\ch (q+\mu) } 
\\&=e^{-2 \pi i \zeta \mu} \frac{e^{2 \pi i \mu p }}{\ch (p - \zeta')} 
\left( e^{\pi p} \frac{e^{2\pi i \zeta q} }{\ch q} e^{\pi p} - 
e^{-\pi p} \frac{e^{2\pi i \zeta q} }{\ch q} e^{-\pi p} \right)
\frac{e^{-2 \pi i \mu p }}{\ch (p + \zeta')}\,,
\eal
to bring the density operator into the form 
\bal \rho &=\frac{1}{4} e^{- 2\pi i (\zeta^{(1)}+\zeta^{(2)}+\zeta^{\prime(2)} ) \mu } 
e^{- \pi i \zeta^{(1)} q} e^{2\pi i \mu p } 
\\&\quad{}\times
\frac{e^{-2\pi i \mu p }}{\ch \big(p + \tfrac{1}{2} \zeta^{(1)} +\zeta^{\prime(2)}\big) } 
\frac{1 }{\ch \big(p + \tfrac{1}{2} \zeta^{(1)}\big) } 
\frac{1}{\ch \big(p - \tfrac{1}{2} \zeta^{(1)}\big) } 
\frac{1}{\ch \big(p - \zeta^{\prime(0)}- \tfrac{1}{2} \zeta^{(1)}\big) }
\\&\quad{}\times
\left( e^{\pi p} \frac{e^{2\pi i (\zeta^{(0)} + \zeta^{\prime(0)} +\zeta^{(1)} )q} }{\ch q} e^{\pi p}
+e^{-\pi p} \frac{e^{2\pi i (\zeta^{(0)} + \zeta^{\prime(0)} +\zeta^{(1)} )q} }{\ch q} e^{-\pi p} \right)
\\&\quad{}\times
\frac{e^{2\pi i \mu p }}{\ch \big(p + \zeta^{\prime(0)}+ \tfrac{1}{2} \zeta^{(1)}\big) }
\frac{1}{\ch \big(p + \tfrac{1}{2} \zeta^{(1)}\big) }
\frac{1}{\ch \big(p - \tfrac{1}{2} \zeta^{(1)}\big) }
\frac{1}{\ch \big(p - \tfrac{1}{2} \zeta^{(1)} - \zeta^{\prime(2)}\big) }
\\&\quad{}\times
\left( e^{\pi p}
\frac{e^{2\pi i (\zeta^{(2)} + \zeta^{\prime(2)} +\zeta^{(1)} )q} }{\ch q} e^{\pi p} + e^{-\pi p}
\frac{e^{2\pi i (\zeta^{(2)} + \zeta^{\prime(2)} +\zeta^{(1)} )q} }{\ch q} 
e^{-\pi p} \right)e^{-2\pi i \mu p } e^{\pi i \zeta^{(1)} q}\,.
\eal 

Applying the canonical transformation \eqref{mirrortrans} 
and removing the exponential factors at the beginning and end by conjugation, 
we obtain (up to an overall phase)%
\footnote{Such phases are unphysical, see footnote~\ref{phase}.}
the density 
operator of the (second) mirror dual theory, which is a $U(2N)$ theory with two anti-symmetric 
hypermultiplets of masses $\widetilde{M}_1,\widetilde{M}_2$, four fundamental hypermultiplets 
of masses $\tilde\mu_\alpha$, $\alpha =1, \cdots, 4$ and an FI parameter $\tilde\zeta$. 
The explicit mirror map between parameters is given by
\bal
\widetilde{M}_1 &= -\zeta^{(0)} - \zeta^{\prime(0)} - \zeta^{(1)}, \qquad 
\widetilde{M}_2 = -\zeta^{(2)} - \zeta^{\prime(2)} - \zeta^{(1)},\\
\tilde\mu_1 &= \tfrac{1}{2} \zeta^{(1)} +\zeta^{\prime(0)}, \qquad 
\tilde\mu_2 = \tfrac{1}{2} \zeta^{(1)}, \qquad
\tilde\mu_3 = -\tfrac{1}{2} \zeta^{(1)}, \qquad 
\tilde\mu_4 = -\tfrac{1}{2} \zeta^{(1)} - \zeta^{\prime(2)},\\
\tilde\zeta &= \mu\,. 
\eal

Finally we may think of other ways to pair the $U(N)$ nodes of the $\hat D_4$-quiver which 
are similar to the two cases above. For instance we can consider the exchange 
$\zeta^{\prime(0)} \leftrightarrow \zeta^{\prime(2)}$ in \eqn{D4rhonoz2}.
This symmetry, which is completely trivial on the $\hat{D}$-quiver side, manifests as a 
relation between two mirror $U(2N)$ theories which differ only by the values of their 
mass parameters
\bal
\label{mirroredsymmetry}
\widetilde M_1 &\rightarrow \widetilde M_1 + \tilde\mu_1 + \tilde\mu_4\,, \qquad
\widetilde M_2 \rightarrow \widetilde M_2 - \tilde\mu_1 - \tilde\mu_4\,, \\
\tilde\mu_1 &\rightarrow -\tilde\mu_4, \qquad \tilde\mu_2 \rightarrow \tilde\mu_2, 
\qquad \tilde\mu_3 \rightarrow \tilde\mu_3,\qquad \tilde\mu_4 \rightarrow -\tilde\mu_1\,, \\
\tilde\zeta &\rightarrow \tilde\zeta\,. 
\eal

\subsection{$\hat{D}_5$-quiver}

\begin{figure}[ht]
\centering
\epsfig{file=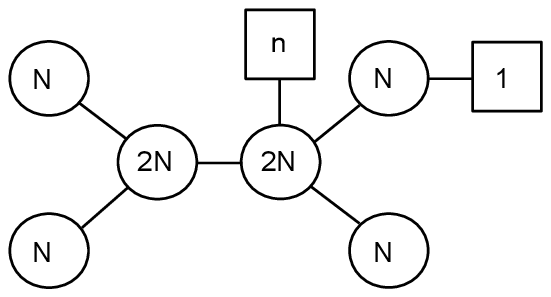,width=0.47\textwidth
\psfrag{n}{$n$}
\psfrag{N}{$N$}
\psfrag{2N}{$2N$}
\psfrag{1}{$1$}
}\hspace{0.3cm}
\epsfig{file=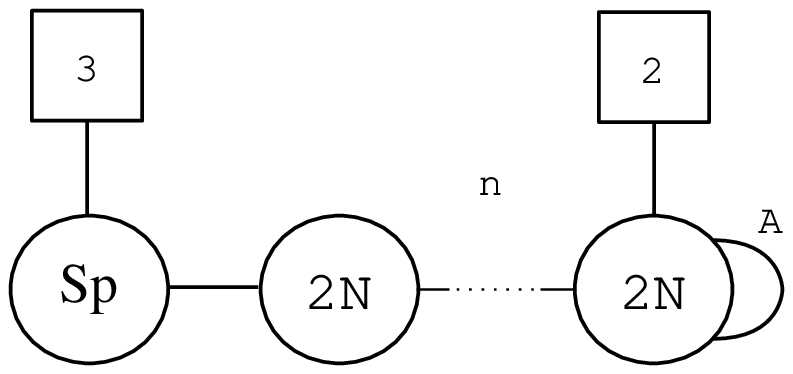,width=0.47\textwidth
\psfrag{Sp}{$\hspace{-0.4cm} Sp(2N)$}
\psfrag{2N}{$2N$}
\psfrag{3}{$3$}
\psfrag{2}{$ 2$}
\psfrag{A}{$A$}
\psfrag{n}{$\hspace{-1.4cm} { \overbrace{\hspace{2.8cm}}^{\displaystyle n} }$}
}
\caption{$\hat D_5$-quiver and its mirror dual theory. }
\label{fig:D5}
\end{figure}

The next example of mirror map involves 
a $\hat{D}_5$-quiver with $n$ fundamental hypermultiplets on one $U(2N)$ node 
and a single fundamental hypermultiplet on one $U(N)$ node. The quiver is shown in 
figure~\ref{fig:D5}. Shifting eigenvalues to set all bifundamental masses, 
and one of the $U(2N)$ fundamental masses to zero, 
we find the density operator (with obvious notations for the mass parameters)
\bal
\label{rhoD5}
\rho &= \frac{1}{4}\frac{1}{\ch p} \left( e^{2 \pi i \zeta^{(0)}q } 
\frac{\sh p}{\ch p} e^{2 \pi i \zeta^{\prime(0)}q } + e^{2 \pi i \zeta^{\prime(0)}q }
\frac{\sh p}{\ch p} e^{2 \pi i \zeta^{(0)} q } \right) \frac{1}{\ch p} e^{2 \pi i \zeta^{(1)} q} 
\\&\quad{}\times 
\frac{1}{\ch p}\frac{ e^{2 \pi i \zeta^{(2)} q}}{\ch q \prod_{\alpha=1}^{n-1} \ch \big(q + \mu^{(2)}_\alpha \big) } 
\frac{1}{\ch p}
\left( \frac{e^{2 \pi i \zeta^{(3)}q } }{\ch \big( q + \mu_{\phantom{1}}^{(3)} \big) } 
\frac{\sh p}{\ch p} e^{2 \pi i \zeta^{\prime (3)}q } 
\right.\\&\qquad\left.{}
+ e^{2 \pi i \zeta^{\prime (3)}q }
\frac{\sh p}{\ch p} \frac{e^{2 \pi i \zeta^{(3)} q } }{\ch \big( q + \mu_{\phantom{1}}^{(3)} \big) } \right) 
\frac{1}{\ch p} 
\frac{ e^{2 \pi i \zeta^{(2)} q}}{\ch q \prod_{\alpha =1}^{n-1} \ch \big(q + \mu^{(2)}_\alpha \big) } 
\frac{1}{\ch p}e^{2 \pi i \zeta^{(1)} q} \, .
\eal

Using the identities \eqn{shiftid}, \eqn{mirrorid1} and 
\eqn{mirrorid2}, we can write the density operator as
\begin{align}
\rho &= \frac{1}{2}e^{-2 \pi i(\zeta^{(3)} + \zeta^{\prime(3)})\mu^{(3)}_{\phantom{1}} }\frac{1}{\ch p} 
e^{ \pi i ( \zeta^{(0)} + \zeta^{\prime(0)})q } 
\frac{\sh 2 p}{\ch (p+ \tfrac{1}{2} \zeta^{(0)} - \tfrac{1}{2} \zeta^{\prime(0)})
\ch (p - \tfrac{1}{2} \zeta^{(0)} + \tfrac{1}{2} \zeta^{\prime(0)}) }
\nonumber\\
&\quad{}\times
e^{ \pi i ( \zeta^{(0)} + \zeta^{\prime(0)})q } \frac{1}{\ch p} e^{2 \pi i \zeta^{(1)} q} \frac{1}{\ch p} 
\frac{ e^{2 \pi i \zeta^{(2)} q}}{\ch q \prod_{\alpha=1}^{n-1} \ch \big(q + \mu^{(2)}_\alpha \big) } \frac{1}{\ch p} 
\nonumber\\
&\quad{}\times
\frac{e^{2 \pi i \mu^{(3)}_{\phantom{1}} p }}{\ch (p - \zeta^{\prime(3)}) } 
\left(e^{\pi p} 
\frac{e^{2\pi i (\zeta^{(3)} + \zeta^{\prime(3)})q} }{\ch q}e^{\pi p} 
+ e^{-\pi p} \frac{e^{2\pi i (\zeta^{(3)} + \zeta^{\prime(3)})q} }{\ch q} 
e^{-\pi p} \right)
\frac{e^{-2 \pi i \mu^{(3)}_{\phantom{1}} p }}{\ch (p + \zeta^{\prime(3)})} 
\nonumber\\
&\quad{}\times
\frac{1}{\ch p} \frac{ e^{2 \pi i \zeta^{(2)} q}}{\ch q \prod_{\alpha=1}^{n-1} \ch \big(q + \mu^{(2)}_\alpha \big) } 
\frac{1}{\ch p} e^{2 \pi i \zeta^{(1)} q} \, .
\label{rhoD5trans}
\end{align}
Further commuting exponential terms using \eqn{shiftid} we obtain
\bal
\label{rhoD5final}
\rho = \frac{1}{2} & 
e^{\pi i (\zeta^{(0)}+\zeta^{\prime(0)}) q}\ch \big(p + \tilde\mu^{(0)}_3 \big) \ch q \, 
e^{2 \pi i \widetilde m q } \, 
e^{-2 \pi i (\zeta^{(3)} + \zeta^{\prime(3)}) \mu^{(3)}} \\
\times &\frac{e^{-2 \pi i \widetilde m q }}{\ch q} 
\frac{\sh 2 p}{ \prod_{\beta =1}^3 \ch \big(p + \tilde\mu^{(0)}_\beta \big) \ch \big(p - \tilde\mu^{(0)}_\beta \big) } 
\frac{e^{-2 \pi i \widetilde m q }}{\ch q} 
\left( \prod_{a=1}^{n-1} e^{2 \pi i \tilde\zeta^{(a)}p } \frac{1}{\ch q} \right) \\
\times &
\frac{e^{2 \pi i \tilde\zeta^{(n)} p } }{\ch p \ch \big(p + \tilde\mu_{\phantom{1}}^{(n)}\big) } 
\left( e^{\pi p} \frac{e^{-2 \pi i \widetilde M q}}{\ch q } e^{\pi p}
+ e^{-\pi p} \frac{e^{-2 \pi i \widetilde M q}}{\ch q } e^{-\pi p} \right)
\frac{e^{-2 \pi i \tilde\zeta^{(n)} p } }{\ch p \ch \big(p- \tilde\mu_{\phantom{1}}^{(n)}\big) } \\
\times & \left( \prod_{a=1}^{n-1} \frac{1}{\ch q} e^{-2 \pi i \tilde\zeta^{(n-a)}p } \right)
\frac{e^{-2 \pi i \widetilde m q }}{\ch q} \frac{1}{\ch \big(p + \tilde\mu^{(0)}_3 \big)} 
e^{-\pi i (\zeta^{(0)}+\zeta^{\prime(0)}) q} ,
\eal
with 
\bal 
\widetilde m &= - \tfrac{1}{2} \zeta^{(0)} - \tfrac{1}{2} \zeta^{\prime(0)} - \zeta^{(1)} - \zeta^{(2)},\qquad 
\widetilde M = -\zeta^{(3)} - \zeta^{\prime(3)} \\
\tilde\mu^{(0)}_1 &= \tfrac{1}{2}\zeta^{(0)} -\tfrac{1}{2}\zeta^{\prime(0)}, \qquad 
\tilde\mu^{(0)}_2 = \tfrac{1}{2}\zeta^{(0)} +\tfrac{1}{2}\zeta^{\prime(0)}, \qquad
\tilde\mu^{(0)}_3 = \tfrac{1}{2}\zeta^{(0)} +\tfrac{1}{2}\zeta^{\prime(0)} + \zeta^{(1)}, \\
\tilde\mu_{\phantom{1}}^{(n)} &= -\zeta^{\prime(3)}, \\
\tilde\zeta^{(1)} &= \mu^{(2)}_1, \qquad 
\tilde\zeta^{(n)} = \mu^{(3)}_{\phantom{1}} - \mu^{(2)}_{n-1}, \qquad
\tilde\zeta^{(a)} = \mu^{(2)}_a -\mu^{(2)}_{a-1}, \quad a= 2,\cdots, n-1\,.
\label{D5mirrormap}
\eal

Applying the canonical transformation \eqref{mirrortrans} and removing terms 
at the beginning and end of \eqn{rhoD5final} by conjugation, we obtain 
up to an overall phase (see footnote~\ref{phase}) the density operator
of the mirror theory. This is a linear quiver with one $Sp(2N)$ node with three 
fundamental hypermultiplets, connected to $n$ $U(2N)$ nodes, where the last $U(2N)$ node 
has one antisymmetric and two fundamental hypermultiplets. 
By shifting eigenvalues 
the masses of hypermultiplets transforming in the bifundamental representation of two $U(2N)$ 
nodes and the mass of one of the $U(2N)$ fundamental hypermultiplets can be set to zero. 
This leaves us with the masses: $\widetilde m$ for the $Sp(2N)\times U(2N)$ bifundamental 
hypermultiplet, $\widetilde M$ for the antisymmetric hypermultiplet, $\tilde\mu^{(0)}_{1,2,3}$ 
for the three $Sp(2N)$ fundamental hypermultiplets, $\tilde\mu_{\phantom{1}}^{(n)} $ for one 
$U(2N)$ fundamental hypermultiplet. Moreover the theory has FI parameters 
$\tilde \zeta^{(a)}$, for $a= 1,\cdots, n$. The mirror map between parameters 
is given by \eqref{D5mirrormap}.

This general approach to finding mirror maps can be easily extended to 
more general $\hat{D}$-quivers. 
For mirrors involving a `bad' linear quiver, proposed in \cite{Dey2014a} 
the matrix model of the linear 
quiver is divergent,%
\footnote{It was proposed in \cite{Kapustin2010a} that the divergence of `bad' 
quiver matrix model is related to a mismatch between the R-symmetry group in the UV localization
computation, and the R-symmetry group at the infrared fixed point.} 
while the matrix model of the $\hat D$ quiver is still finite. 
As we mention in the discussion section, the density operators of the `bad' linear 
quivers can be matched to those of the `good' $\hat D$-quivers with a simple replacement, 
alas, this replacement is rather ad hoc and we do not know whether it represents a true 
regularization of the non trace class density operator of the `bad' quiver.

\section{Computing the partition function}
\label{sec:partition}

Having re-expressed the $S^3$ partition function of $\hat D$-quiver theories as a free fermion
partition function, we proceed now with its evaluation following the technique developed in 
\cite{Marino2012}. More precisely we compute the perturbative part in the large $\mu$ 
expansion of the grand potential $J$ 
and extract from it the perturbative part in the large~$N$ expansion of the 
partition function $Z$, which turns out to be the Airy function \eqref{ZNAiryintro}. 
Our strategy to compute $J$ has new ingredients compared to \cite{Marino2012}, 
in particular we use a simplifying recursion method to compute the perturbative expansion of 
$\Tr \rho^l$, for arbitrary $l\ge 1$. This method can be used for any density operator $\rho$.
We also rely on the integral representation of $J$, following \cite{Hatsuda2015}, to evaluate 
its perturbative part.
Masses and FI parameters introduce extra difficulties in the computation and we set
all of them to zero in this section.

We proceed to present this strategy and provide the exact evaluations for all the 
$\hat D$-quivers considered in this paper.

\subsection{General strategy}

The starting point for our computation of the perturbative part of the $\hat D$-quiver
(and linear mirror)
partition functions \eqn{ZNgeneral} is the Fermi gas reformulation
\bal
\label{ZNgeneral2}
Z(N) &= \frac{1}{N!}\sum_{\sigma \in S_N} \frac{(-1)^\sigma }{2^{n_\sigma}} \int \prod_{i=1}^N d\lambda_i \,
\prod_{i=1}^N \rho(\lambda_i, \lambda_{\sigma(i)})\,,
\eal
where we recall that $n_\sigma$ counts the number of cycles in $\sigma$. 
The standard analysis \cite{Feynman1972} is to factor the integral into its closed loops
\beq
\label{loops}
\int d \lambda_1 \cdots d \lambda_l \, \rho(\lambda_1, \lambda_2 ) \rho(\lambda_2, \lambda_3 ) 
\cdots \rho(\lambda_l, \lambda_1 ) 
= \Tr \rho^l \equiv Z_l\,.
\eeq
These loops of course correspond to the cycles of the permutation $\sigma$, and so the summand 
of \eqn{ZNgeneral2} depends only on $Z_l$ and the conjugacy class of $\sigma$. Conjugacy
classes of $S_N$ can be labelled a set of integers $\{ m_l \}$, where $m_l$ 
is the number of cycles of length $l$. In terms of this labelling we have
\beq
\frac{1}{2^{n_\sigma}} = \prod_l \frac{1}{2^{m_l}}\,,
\eeq
and the number of permutations in a given conjugacy class is given by
\beq
\label{SNcombinatorics}
\frac{N !}{\prod_l m_l! \, l^{m_l}}\,.
\eeq
With these combinatorics \eqn{ZNgeneral2} becomes
\beq
\label{ZNZl}
Z(N) = {\sum_{\{m_l\}}}^\prime \prod_l \frac{ (\tfrac{1}{2}Z_l)^{m_l} \, (-1)^{(l-1)m_l}} {m_l!\, l^{m_l} }\,,
\eeq
where the primed sum denotes a sum over sets that satisfy $\prod_l l m_l = N $.

The computation of $Z(N)$ thus boils down to the evaluation
of $Z_l$. 
To compute results for large~$N$, the standard approach\cite{Feynman1972} is consider the grand 
canonical partition function 
\beq
\Xi(z) = 1 + \sum_{N=1}^\infty Z(N) z^N = e^{J(\mu)}, \qquad z= e^\mu\,, 
\eeq
where $\mu$ is the chemical potential, and $J(\mu)$ the grand canonical potential, given by
\beq
\label{Jmu}
J(\mu) = -\sum_{l= 1}^\infty \frac{ (-1)^l Z_l e^{ \mu l} }{2 l}\,.
\eeq
The strategy is to first find an expression for $Z_l$, then to resum the expression and obtain 
$J(\mu)$ using \eqn{Jmu}, and finally to recover $Z(N)$ by computing
\beq
\label{ZNJmu}
Z(N) = \frac{1}{2 \pi i} \int_{\mu_0 - \pi i}^{\mu_0 + \pi i} d \mu \, e^{J(\mu) - \mu N}\,,
\eeq
where $\mu_0$ can be chosen arbitrarily without affecting the result.

In practice, computing $Z_l$ exactly for arbitrary $l$ is 
highly non trivial. To make the problem tractable it is useful to reformulate it within 
Wigner's phase space \cite{Marino2012}. For a general review of Wigner's phase space 
see \cite{Zachos2005}; here we simply summarise the properties
that we require. 

The Wigner transform of an operator%
\footnote{To avoid confusion between phase space variables $p,q$, and the canonical position and 
momentum operators, we give all operators hats for the remainder of this section. } 
$\hat A$ is given (with $\hbar=\frac{1}{2\pi}$) by
\beq
\label{wigtrans}
A_W(q,p) = \int d q^\prime 
\Bra{q - \frac{q^\prime}{2}}\hat A \Ket{q + \frac{q^\prime}{2}}e^{2 \pi i p q^\prime }\,.
\eeq
Some important identities are 
\beq
\label{wigident}
(\hat A \hat B)_W = A_W \star B_W, \qquad \star 
= \exp \left[ \frac{i }{4 \pi}\left(\overset{\leftarrow}\partial_q\overset{\rightarrow}\partial_p
- \overset{\rightarrow}\partial_q\overset{\leftarrow}\partial_p\right) \right],
\qquad
\Tr(\hat A) = \int dp dq A_W\,.
\eeq
In the language of phase space $Z_l$ becomes
\beq
\label{Zlwig}
Z_l = \int d p d q \, \overbrace{\rho_W \star \rho_W \cdots \star \rho_W}^l\,.
\eeq
We generate an expansion for the integrand of \eqn{Zlwig} by performing a derivative 
expansion of the star products \eqn{wigident}.
To this end, we introduce into the star product an expansion parameter 
$\epsilon$ which will be set at the end to 1.%
\footnote{In addition to the differential expression for the star product \eqn{epsilonstarexpand}, 
it also has an equivalent integral representation
\beq
\label{starintid}
(f \star g) (p,q) = \frac{4}{\epsilon^2} \int dq' dp' d q'' d p'' f(p+p', q+q') g(p+p'', q+q'') 
e^{4\pi i/\epsilon (q' p'' - q''p') }\,.
\eeq
In some cases (for instance when $f$ or $g$ involve delta functions) \eqn{starintid} produces an 
exact result that is non perturbative in $\epsilon$ \cite{Bars2001}. In these cases extra care should be 
taken, because the perturbative star product \eqn{wigident} is not valid.}
\bal
\label{epsilonstarexpand}
\star 
&= \exp \left[ \frac{i \epsilon }{4 \pi}\left(\overset{\leftarrow}\partial_q\overset{\rightarrow}\partial_p
- \overset{\rightarrow}\partial_q\overset{\leftarrow}\partial_p\right) \right] 
\\&= 1 + 
\frac{i \epsilon}{4 \pi}\left(\overset{\leftarrow}\partial_q\overset{\rightarrow}\partial_p
- \overset{\rightarrow}\partial_q\overset{\leftarrow}\partial_p\right)
-\frac{ \epsilon^2}{32 \pi^2}\left(\overset{\leftarrow}{\partial^2_q}\overset{\rightarrow}{\partial^2_p}
+ \overset{\rightarrow}{\partial^2_q} \overset{\leftarrow}{\partial^2_p} 
-2\overset{\leftarrow}\partial_{q,p} \overset{\rightarrow}\partial_{q,p} \right)
+\cO (\epsilon^3)\,.
\eal 

In \cite{Marino2012}, the role of the expansion parameter $\epsilon$ was played by the
Planck constant $\hbar$, which was proportional to the Chern-Simons level of the ABJM theory.
We do not have a tunable $\hbar$ here, however it still proves useful to consider the derivative 
expansion associated to $\epsilon$, as we now explain.

The $\epsilon$ expansion of the integrand of \eqn{Zlwig} takes the form 
\beq
\label{rholhbarexpand}
(\hat\rho^l)_W (p,q) = \sum_{n\geq 0} \epsilon^{ n} \,\rho_{l(n)}(p,q).
\eeq
Note that the $\epsilon$ factors come from the expansion of the star products present 
in \eqref{Zlwig}, as well as those arising when replacing the density operator \eqref{rhoDgen}
by its Wigner transform.
$Z_l$ can then be evaluated order by order in $\epsilon$
\beq
\label{Zlexpand}
Z_l = \sum_{n \geq 0} \epsilon^{n} Z_{l(n)}, \qquad Z_{l(n)} = \int d p d q \, \rho_{l(n)}\,.
\eeq
Resumming each term using \eqn{Jmu} then generates an $\epsilon$ expansion for $J(\mu)$ 
\beq
\label{Jmuexpand}
J(\mu) = \sum_{n \geq 0} \epsilon^{n} J_{(n)}(\mu), \qquad 
J_{(n)}(\mu)= -\sum_{l= 1}^\infty \frac{ (-1)^l Z_{l(n)} e^{ \mu l} }{2 l}\,.
\eeq

As for the $\hat A$-quivers \cite{Marino2012}, 
we anticipate that
for $\hat D$-quivers (and their linear mirrors), $J(\mu)$ also admits an asymptotic expansion 
of the form
\beq
\label{Jmuasymp}
J(\mu) = \frac{C(\epsilon )}{3} \mu^3 + B(\epsilon) \mu + A(\epsilon)
+ \cO (e^{-\alpha \mu})\,,
\qquad 
\alpha>0\,,
\eeq
where each of the coefficients $A$, $B$ and $C$ are given by power expansions in 
$\epsilon$. In principle, to obtain a meaningful result for $A$, $B$ and $C$ we should now 
compute and then resum an infinite series of corrections in powers of $\epsilon$, which are 
really all of the same order since $\epsilon = 1$. 
From the study of $\hat A$-quivers, it is expected that the expansions of 
$C(\epsilon)$ and $B(\epsilon)$ truncate at orders $\epsilon^0$ and $\epsilon^2$ 
respectively, so that the first few orders of the $\epsilon$ expansion
are sufficient to compute them exactly. 
We give a proof of this truncation for $\hat D$-quivers with equal number of fundamental 
hypermultiplets on each pair of terminating $U(N)$ nodes in
appendix~\ref{sec:epsilontruncation}. We assume that it holds for the other quivers as well.%
\footnote{We checked that $C$ and $B$ do not receive contributions at order 
$\epsilon^4$ for all $\hat D$-quivers.}

It remains to plug the result \eqn{Jmuasymp} into \eqn{ZNJmu} to extract the perturbative
part of $Z$ at large~$N$. 
In practice the evaluation is done by setting the contour integral parameter
$\mu_0$ to the saddle point $\mu^{\ast}$ of the integrand and by extending the contour along all 
the imaginary axis to 
$(\mu^{\ast} -i\infty, \mu^{\ast}+ i \infty)$. 
As explained in \cite{Marino2012,Hatsuda2013}, this change of contour does not affect the 
perturbative part of the result. The integration leads to the Airy 
function behaviour of the partition function at large~$N$, which is our main result
\beq
\label{ZNAiry}
Z(N) = C^{-\frac{1}{3} } e^A \Ai\left[C^{-\frac{1}{3}}(N -B ) \right] 
+ Z_{\text{np}}(N)\,, 
\eeq
where $ Z_{\text{np}}(N)$ denotes non-perturbative, exponentially suppressed contributions, 
and we note that the undetermined coefficient $A$ only affects the overall prefactor.

\subsection{Recursive formula for $(\hat \rho^l)_W$}

In this subsection we present a simple recursive approach 
for evaluating the coefficients in the $\epsilon$ expansion
of $(\hat \rho^l)_W$ \eqn{rholhbarexpand}. 
This comes from the $\epsilon$ expansion of
\beq
\label{recursion}
(\hat\rho^{l+1})_W = (\hat\rho^{l})_W \star \rho_W\,.
\eeq

One first needs to evaluate the $\epsilon$ expansion of $\rho_W$ 
(which is due to replacing all operator products by star products)
\beq
\label{rhohbarexpand}
\rho_W(p,q) = \sum_{n\geq 0} \epsilon^{ n} \,\rho_{(n)} (p,q)\,,
\eeq
which also serves as the initial conditions for the recursion $\rho_{1(n)}=\rho_{(n)}$.

At order $\epsilon^0$, equation \eqn{recursion} gives
\beq
\rho_{l+1(0)}^{} = \rho_{l(0)}^{} \rho_{(0)}^{}\,,
\eeq
which is solved by
\beq
\label{leading recursion} 
\rho_{l(0)}^{} = \rho_{(0)}^l\,.
\eeq
At order $\epsilon^1$ we get
\beq
\rho_{l+1(1) }^{} = \rho_{(0)}^l \rho_{(1)}^{} + \rho_{(0)}^{} \rho_{l(1)}^{} 
\qquad\Rightarrow\qquad
\rho_{l(1) }^{} = l \rho_{(0)}^{\smash{l-1} } \rho_{(1)}^{}\,. 
\eeq
At order $\epsilon^2$ we then get
\bal
\rho_{l+1(2) }^{} 
&= \rho_{(0)}^{} \rho_{l(2)}^{}
+ \rho_{(0)}^l \rho_{(2)}^{} 
+l \rho_{(0)}^{\smash{l-1}} \rho_{(1)}^2 
-\frac{1}{32 \pi^2}\rho_{(0)}^{l-2} l
\Big[2\big(\rho_{(0)}^{} \rho_{(0)}'' \ddot{\rho}_{(0)}^{} -\rho_{(0)}^{}
\dot{\rho}_{(0)}^{\prime 2} \big)
\\&\hskip2in{}
+(l-1) 
\big( \ddot{\rho}_{(0)}^{} \rho_{(0)}^{\prime 2} + \rho_{(0)}'' 
\dot{\rho}_{(0)}^2 - 2 \rho_{(0)}' \dot{\rho}_{(0)}^{} \dot{\rho}_{(0)}' \big)\Big], 
\eal
where 
\beq
\dot{f}(p,q) \equiv \partial_p \, f(p,q)\,, 
\qquad f'(p,q) \equiv \partial_q \,f(p,q)\,. 
\eeq
Solving the recurrence relation, with initial condition $\rho_{1(2)} = \rho_{(2)}$ yields
\bal
\label{coeff2}
\rho_{ l(2)} = l \rho_{(0)}^{\smash{l-1}} \rho_{(2)} + \frac{1}{2} l(l-1) \rho_{(0)}^{\smash{l-2}} 
\rho_{(1)}^2 - \frac{1}{96 \pi^2} \rho_{(0)}^{l-3} l(l-1) 
\Big[ &3\big(\rho_{(0)}^{} \rho_{(0)}'' \ddot{ \rho}_{(0)}^{} -\rho_{(0)}^{} 
\dot{\rho}_{(0)}^{\prime 2} \big)
\\&\hskip-3.4cm{}
+(l-2) \big( \ddot{\rho}_{(0)}^{} \rho_{(0)}^{ \prime 2 } + \rho_{(0)}'' 
\dot{\rho}_{(0)}^2 - 2 \rho_{(0)}' \dot{\rho}_{(0)}^{} \dot{\rho}_{(0)}' \big)\Big].
\eal
This procedure can be straightforwardly continued to higher order in $\epsilon$.
Finally we plug the expansion coefficients into \eqn{Zlexpand} to obtain the 
$\epsilon$ expansion of $Z_l$. In particular this gives 
\bal
\label{Zlbyparts} 
Z_{l(0)} &= \int dp dq \, \rho_{(0)}^l\,,\\
Z_{l(1)} &= \int dp dq \, l \rho_{(0)}^{\smash{l-1} } \rho_{(1)}^{}\,,\\
Z_{l(2)} &= \int dp dq \, \left( l \rho_{(0)}^{\smash{l-1}} \rho_{(2)}^{}
+ \frac{1}{2} l(l-1) \rho_{(0)}^{\smash{l-2}} \rho_{(1)}^2 
-\frac{1}{96 \pi^2} l^2 (l-1)(l-2) \rho_{(0)}^{\smash{l-4}} \dot{\rho}_{(0)}^2 \rho_{(0)}^{\prime 2} \right).
\eal
Where the last line follows from integrating \eqn{coeff2} by parts, and using the fact that for $\hat A$ or
$\hat{D}$ quivers we always have the decomposition $\rho_{(0)}^{} =t(p)u(q)$. 

We stress that this algorithm is very general and can be applied to any $\hat A$ or $\hat D$-quiver.
All one needs in order to compute $Z_l$ is to plug into $\eqn{Zlbyparts}$ 
the $\epsilon$ expansion of the density operator itself.

\subsection{Computing $Z(N)$ for $\hat D $-quivers}

We now show how to apply the approach outlined in the previous sections and compute 
$Z(N)$ for a generic $\hat D$-quiver with arbitrary number of nodes 
and arbitrary number of fundamental hypermultiplets on each node.
We set here all masses and FI parameters to zero, as this simplifies 
the explicit evaluation of the phase space integrals.
The quiver diagram is shown in figure~\ref{fig:Dgen}, and the density operator is given by 
\eqn{rhoDgen} with
\bal
F^{(a)}(q) = \frac{1}{\ch^{n^{(a)} } q}, \qquad F^{\prime(a)}(q) = \frac{1}{\ch^{n^{\prime(a)} } q}\,.
\eal

We first work out the $\epsilon$-expansion of $\rho_W$ itself, which is then plugged in to 
the result of the recursive formula \eqref{Zlbyparts}.

Using manipulations similar to those used in section~\ref{sec:mirror}
\begin{align}
&\frac{1}{\ch^{n^{(0)} }q }\star \frac{\sh p }{\ch p} \star \frac{1}{\ch^{n^{\prime(0)} }q }
+\frac{1}{\ch^{n^{\prime(0)} }q} \star \frac{\sh p }{\ch p} \star \frac{1}{\ch^{n^{(0)} }q }
\nonumber\\
&=\frac{1}{\ch^{\min (n^{(0)},n^{\prime(0)} ) }q } \star \frac{1}{\ch p} 
\star \Big( \sh p \star \frac{1}{\ch^{ |n^{(0)} -n^{\prime(0)}|} q}\star \ch p +\ch p 
\star \frac{1}{\ch^{ |n^{(0)} -n^{\prime(0)}|} q}\star \sh p\Big) 
\nonumber\\
& \hspace{0.5cm} \star \frac{1}{\ch p} \star \frac{1}{\ch^{\min (n^{(0)},n^{\prime(0)} ) }q }
\nonumber\\
&=\frac{1}{\ch^{\min (n^{(0)},n^{\prime(0)} ) }q } \star \frac{1}{\ch p} 
\star \frac{2 \sh 2 p}{\ch^{ |n^{(0)} -n^{\prime(0)}|} q} \star \frac{1}{\ch p} 
\star \frac{1}{\ch^{\min (n^{(0)},n^{\prime(0)} ) }q }\,,
\label{rhomanip}
\end{align}
where in the last step we evaluated the expression inside parentheses using
the exact star product \eqn{starintid}. 
This allows us to write the Wigner transform of the density operator \eqn{rhoDgen} as%
\footnote{$\prod \mathstrut{\vphantom\prod}_\star$ is defined by ordered star multiplication}
\bal
\rho_W = &\, e^{T(p)} \star e^{U_1(q) } \star e^{T(p)} \star e^{S (p) +2 U_0 (q)} 
\star e^{T(p)} \star \bigg( \prod_{k=1}^{\lambda-1}\mathstrut{\vphantom\prod}_\star 
e^{U_{k} (q) } \star e^{T(p)} \bigg) \star e^{S (p) +2 U_\lambda(q)} 
\\&\star \bigg( \prod_{k=1}^{\lambda-2}\mathstrut{\vphantom\prod}_\star e^{T(p)} \star e^{U_{\lambda-k} (q) } \bigg)\,,
\eal
where
\bal
\label{UTdefs}
S (p) = \log \sh 2p, \qquad T(p) = \log \frac{1}{\ch p}, \qquad U_i(q) = 
\log \frac{1}{\ch^{\eta_i } q}\,, 
\eal
and 
\bal
&\eta_{0} = \frac{1}{2} |n^{(0)} - n^{\prime (0)} |\,,
\qquad 
\eta_{1} = \min( n^{(0)}, n^{\prime(0)})\,,
\qquad
\eta_{i} = n^{(i-1)}\,,\quad i= 2, \cdots, L
\\&\eta_{\lambda} = \frac{1}{2} |n^{(L)} - n^{\prime (L)} |\,, 
\qquad 
\eta_{\lambda-1} =\min( n^{(L)}, n^{\prime(L)})\,,
\qquad
\lambda = L+2\,.
\eal

A first useful manipulation is to conjugate%
\footnote{Since we are ultimately computing $Z_l = \Tr \hat\rho^l$ we can use cyclicity 
of the trace to make conjugations $\hat \rho \rightarrow \hat V^{-1} \hat \rho \hat V$, which in 
the language of phase space becomes $\rho_W \rightarrow (\hat V^{-1})_W \star \rho_W \star V_W$ }
the density operator into a more symmetric form that resembles a palindrome
\bal
\label{rhoDgenwig}
\rho_W \approx \sqrt[\leftroot{-2}\uproot{2}\star]{e^{S (p) +2 U_\lambda (q)}} \star
e^{T(p)} &\star 
\bigg( \prod_{k=1}^{\lambda-1}\mathstrut{\vphantom\prod}_\star e^{U_{\lambda-k} (q) } 
\star e^{T(p)}\bigg)\star e^{S (p) +2 U_0 (q) }
\\&\star \bigg( \prod_{k=1}^{\lambda-1}\mathstrut{\vphantom\prod}_\star 
e^{T(p)} \star e^{U_{k} (q) } \bigg) \star e^{T(p)} 
\star\sqrt[\leftroot{-2}\uproot{2}\star]{e^{S (p) +2 U_\lambda (q)}}\,,
\eal
where ${ \sqrt[\leftroot{-2}\uproot{2}\star]{}}$ is the star square root, and its expansion 
gives
\bal
\sqrt[\leftroot{-2}\uproot{2}\star]{e^{S (p) +2 U_\lambda (q)}} = 
e^{\frac{1}{2} S + U_\lambda } 
\left(1 + \frac{\epsilon^2}{128 \pi^2}\big(2 \ddot{S} U_\lambda'' 
+2 \ddot{S} U_\lambda^{\prime 2} + \dot{S}^2 U_\lambda'' \big) 
+ \cO(\epsilon^4) \right).
\eal
The reason for making this conjugation to palindromic form is that the $\epsilon$ expansion 
of \eqn{rhoDgenwig} is purely in even powers of $\epsilon$. 
The easiest way then to compute the 
$\epsilon$ expansion is to build it up from the 
central $e^{S (p) +2 U_0 (q) }$ piece. We first compute
\bal
e^T \star e^{S +2 U_0 } \star e^T = e^{2 T + S + 2 U_0} \left(1- \frac{\epsilon^2}{8 \pi^2} \ddot{T} 
\big(U_0'' + 2 U_0^{\prime 2} \big) + \cO (\epsilon^4) \right).
\eal
By plugging this result in, we can then easily compute 
\bal
&e^{U_{1}} \star \left(e^T \star e^{S +2 U_0 } \star e^T \right)\star e^{U_{1}}
\\&\hspace{.8cm} = 
e^{2 T + S + 2 U_0+2 U_1} 
\bigg(1-\frac{\epsilon^2}{8 \pi^2} \ddot{T} (U_0'' + 2 U_0^{\prime 2} )
- \frac{\epsilon^2}{16 \pi^2} U_1'' \big( 2\ddot{T} + \ddot{S} + (2\dot{T} + \dot{S} )^2\big) + \cO (\epsilon^4) \bigg).
\eal

Continuing this procedure we find the full expansion of \eqn{rhoDgenwig} up 
to $\cO (\epsilon^2)$ 
\bal
\rho_W &= e^{ 2\lambda T + 2 S + 2\sum_{k=0}^\lambda U_k } \Bigg[ 1- 
\frac{\epsilon^2}{8 \pi^2}\ddot{T} \sum_{k=0}^{\lambda-1} 
\left( \sum_{j=0}^k U_j'' +2 \bigg(\sum_{j=0}^k U_j' \bigg)^2 \right) 
\\&\hskip1cm
-\frac{\epsilon^2}{16 \pi^2} 
\sum_{k=1}^{\lambda} U_k'' \left(\ddot{S} + 2 k \ddot{T} + \big(\dot{S} +2k \dot{T} \big)^2 \right)
\\&\hskip1cm
-\frac{\epsilon^2}{16 \pi^2} \left(\ddot{S} \bigg(\sum_{k=0}^{\lambda-1}U_k'' 
+2 \sum_{k=0}^{\lambda-1}U_k'\sum_{j=0}^{\lambda}U_j' \bigg)
+ U_\lambda'' \dot{S} \big(\dot{S} +2 \lambda \dot{T} \big) \right) 
+ \cO (\epsilon^4) \Bigg].
\eal
The coefficients of this expansion 
are $\rho_{(0)}$ and $\rho_{(2)}$ \eqn{rhohbarexpand}, which 
serve as the seed for the recursion and can be 
plugged directly into \eqn{Zlbyparts}. We find that $Z_{l(0)}$ is given by
\bal
\label{zl0int}
Z_{l(0)} = \int dp dq \, \rho_{(0)}^l = \int dp dq \, 
\frac{\thh^{2l} p}{\ch^{2 l (\lambda-2)} p \ch^{2 l \nu} q }, \qquad \nu = \sum_{k=0}^\lambda \eta_k\,.
\eal

The expression for $Z_{l(2)}$ is considerably more involved, but it can be simplified by 
integrating by parts to remove all double derivatives. 
Integrating by parts the first term in $Z_{l(2)}$ \eqn{Zlbyparts} 
gives
\bal
&\hskip-.5cm
\int dp dq \, l \rho_{(0)}^{\smash{l-1}}\rho_{(2)}^{} 
\\=&\, \frac{1}{4 \pi^2} \int dp dq \, l^2 
e^{2l(\lambda T + S +\sum_{k=0}^\lambda U_k )}
\Bigg[2 \dot{T} \big(\dot{S}+\lambda \dot{T} \big)\sum_{k=0}^{\lambda-1}
\sum_{j=0}^k U_j' \bigg(\sum_{i=0}^k U_i' -l \sum_{i=0}^\lambda U_i' \bigg) 
\\&{}
+\frac{1}{2}
 \sum_{k=1}^\lambda U_k' 
 \big(\dot{S}+2k \dot{T} \big) \big(\dot{S}+2k \dot{T} -2l\big( \dot{S} + \lambda \dot{T} \big)\big)\sum_{j=0}^\lambda U_j' 
\\&{}
- \bigg(\dot{S}\big( \dot{S} + \lambda \dot{T} \big) (l-1) 
\sum_{k=0}^{\lambda-1}U_k'
- \frac{1}{2} \dot{S}\big(\dot{S} +2 \lambda \dot{T}\big)U_\lambda'\bigg)
\sum_{j=0}^{\lambda}U_j' \Bigg]\,.
\eal
The second (non zero) term in $Z_{l(2)}$ gives 
\bal
&\hskip-1cm\int dp dq \,\frac{-1}{96 \pi^2} l^2 (l-1)(l-2) \rho_{(0)}^{\smash{l-4}} 
\dot{\rho}_{(0)}^2 \rho_{(0)}^{\prime 2} 
\\&= \int dp dq \, 
\frac{-1}{6 \pi^2} l^2 (l-1)(l-2) e^{2l(\lambda T + S +\sum_{k=0}^\lambda U_k )}
\bigg(\sum_{j=0}^k \dot U_j\bigg)^2 (\dot{S} + \lambda \dot{T} )^2\,.
\eal
Combining these expressions, substituting \eqn{UTdefs} and simplifying leads finally to
\begin{align}
Z_{l(2)} = &\int dp dq \, l^2 \frac{\pi^2}{24} \frac{\thh^{2l-2} p \thh^{2} q}
{\ch^{2 l (\lambda-2)} p \ch^{2 l \nu} q }\bigg[3(2l-1) \Delta \nu -(4l^2-1)\nu^2
\nonumber\\
&{}+ 2 \thh^2 p \Big((4l^2-1)(\lambda -1) \nu^2 - 3 \Delta \nu (l (\lambda-2)+1 ) - 6 \Sigma_1 \Big)
\label{zl1int}
\\&{}- \thh^4 p \Big( (4l^2-1) (\lambda-1)^2 \nu^2- 3\lambda^2 \nu^2 +3 \Delta \nu(2l(\lambda-1)+1) 
-12 (\Sigma_2-\Sigma_1) \Big) \bigg]\,,
\nonumber
\end{align}
where
\bal
\Delta &= \eta_0 + \eta_\lambda, \qquad 
\Sigma_1 = \sum_{i>j } \eta_i \eta_j (i-j), \\
\Sigma_2 &= \sum_{i>j } \eta_i \eta_j \left((\lambda-i)^2 +j^2 +\lambda^2\right) 
-\sum_{i=0}^\lambda \eta_i^2 i (\lambda-i).
\eal

In order to evaluate the integrals appearing in $Z_{l(0)}$ and $Z_{l(2)}$, we require 
only the identity
\beq 
\label{trigint}
\int dx \frac{\thh^{a} x }{\ch^b x } = \frac{ (1 + e^{i \pi a} ) 
\Gamma\left( \frac{a +1}{2}\right)\Gamma\left( \frac{b}{2}\right)}
{2^{b+1} \pi \Gamma\left( \frac{a+b+1}{2}\right)}, \qquad \text{Re}(a)>-1, \quad \text{Re}(b)>0
\eeq 
Integrating \eqn{zl0int} and \eqn{zl1int} using \eqn{trigint}, simplifying by 
shifting the arguments of Gamma functions and choosing to express the result in terms 
of $L= \lambda-2$ we find
\bal
Z_{l(0)} &= \frac{1}{2^{2l(\nu + L) }\pi^{3/2}} \frac{\Gamma (l+ \tfrac{1}{2} ) 
\Gamma (\nu l)\Gamma(L l )}{\Gamma ( \nu l + \tfrac{1}{2} ) \Gamma ((L+1)l + \tfrac{1}{2} )}\,,\\
Z_{l(2)} &= \frac{ \pi^{1/2} }{3 \cdot 2^{2l(\nu + L) +3}} l^2 F(l) \frac{\Gamma (l+ \tfrac{1}{2} )
\Gamma (\nu l)\Gamma( L l )}{\Gamma ( \nu l + \tfrac{3}{2} )\Gamma ((L+1)l + \tfrac{3}{2} ) }\,,
\eal
where
\bal
\label{Flpoly}
F(l) &=\nu^2 (1 + (L+1)(L+2) ) - 3 \Delta \nu -3 (\Sigma_1 + \Sigma_2) 
\\&\quad{}+l\big(\nu^2 (L+2)(L+3) -6 \Delta \nu (L+1) - 6 (L+1) \Sigma_1 + 6 \Sigma_2
\big) -2l^2\nu^2 L(L+1)\,. 
\eal
Having computed the $\epsilon$ expansion of $Z_l$, we can now compute 
the $\epsilon$ expansion for $J(\mu)$ by resuming each term using \eqn{Jmuexpand}. As 
for the $\hat A$ quivers, this gives some complicated hypergeometric 
functions \cite{Marino2012, Moriyama2014}, 
from which we can then extract the large $\mu$ asymptotic expansion \eqn{Jmuasymp}. 
Ultimately, we are interested only in the perturbative part
of this expansion. A recent paper \cite{Hatsuda2015} pointed out a very elegant 
and simple way to extract this perturbative piece of $J(\mu)$, just by evaluating a single residue 
involving $Z_l$. 

The first step is to write a Mellin-Barnes integral representation for the infinite sum \eqn{Jmuexpand}
\beq
\label{mellinbarnesJmu}
J_{(n)}(\mu)= -\sum_{l= 1}^\infty \frac{ (-1)^l Z_{l(n)} e^{ \mu l} }{2 l} 
= -\int_{c - i \infty}^{c + i \infty} \frac{d l}{4 \pi i} \Gamma(l)\Gamma(-l) Z_{l(n)} e^{l \mu}\,,
\eeq
where $c$ can be chosen arbitrarily in $(0,1)$, and $Z_{l(n)}$ should now be regarded as a 
function of $l$, analytically continued to the complex plane. 
To see how this integral representation reproduces the infinite sum \eqn{mellinbarnesJmu},
note that for $\mu<0$ we can close the contour of integration 
around the region with $\text{Re}(l)>c$. Since $Z_{l(n)}$ itself has no poles in this region, 
the only enclosed poles are the simple poles of $\Gamma(-l)$ for $l = n \in \mathbb{N}^+$. 
Using the fact that
\beq
\mathop{\text{Res}}\limits_{l=n} \Gamma(-l) = \frac{(-1)^n}{n!}\,,
\eeq
we recover \eqn{mellinbarnesJmu}.

Since we are interested in the asymptotic region $\mu \gg0 $, we close the 
contour of integration around the region $\text{Re}(l)< c$. 
In this region there can be poles due to both $Z_{l(n)}$ and $\Gamma(l)$. 
The residue at $l=0$ turns out to be the only 
one giving a contribution that is not exponentially suppressed at large $\mu$. 
Therefore we can immediately evaluate
\bal
J_{(n)}(\mu) = -\frac{1}{2} \mathop{\text{Res}}\limits_{l=0} \Gamma(l)\Gamma(-l) 
Z_{l(n)} e^{l \mu} +\cO (e^{- \alpha \mu } ), \qquad \alpha >0\,.
\eal
Evaluating in this way the perturbative contributions to $J(\mu)$ from $J_{(0)}(\mu) $ and $J_{(2)}(\mu) $, 
we find the asymptotic expansion of the form \eqn{Jmuasymp}, with $C$ and $B$ coefficients 
\bal
\label{DgenJmuresult}
C &= \frac{1}{4 \pi^2 L \nu}, \\
B &= \frac{1 -3 \Delta \nu +2 \nu^2 + L(3+L)(\nu^2 -1) -3(\Sigma_1 + \Sigma_2) }{12 L \nu }\,.
\eal

As explained before, the coefficients $C$ and $B$ do not receive contributions from 
higher order terms in the $\epsilon$ expansion, so that \eqref{DgenJmuresult} provides the
exact result. These coefficients characterize the full perturbative part of $Z(N)$ as an 
Airy function \eqref{ZNAiry}, up to the overall coefficient $A$ which is undetermined by our analysis.

\section{Discussion}
\label{sec:discussion}

One of the main applications of our results concerns holography.
We have found the complete large~$N$ perturbative result for the partition function of 
(good) $\hat D$-quivers as an Airy function \eqref{ZNAiry}. The large~$N$ expansion of 
the free energy starts with the terms
\beq
-\log Z(N) = \frac{2}{3\sqrt C} N^{\frac 32} - \frac{B}{\sqrt C} N^{\frac 12} + \cdots\,.
\label{freeEnergy}
\eeq
The leading term has the famous $N^{3/2}$ behaviour and its coefficient depends only on the 
parameter $C$. This coefficient has a simple geometric interpretation 
when the theory admits an M-theory holographic dual of the form $AdS_4 \times SE_7$, 
where $SE_7$ is a tri-Sasaki-Einstein manifold. Holography predicts the relation 
\cite{Herzog2011}
\beq
C= \frac{6}{\pi^6} {\rm Vol}(SE_7)\,.
\label{Cholo}
\eeq
The $\hat D$-quivers studied in this paper 
(with vanishing masses and FI terms) can be 
engineered as the low-energy limit of a stack of $2N$ M2-branes at 
an orbifold singularity in $\bC^2/\bZ_{2\nu} \times \bC^2/\bD_{L}$ 
\cite{Porrati:1996xi, Dey2012}, where $\bZ_{2\nu}$ and $\bD_{L}$ are discrete subgroups of $SU(2)$ 
of type $A$ and $D$ respectively. The additional data describing the distribution of fundamental
hypermultiplets in the quiver should be encoded in four-form fluxes on vanishing cycles at the orbifold
singularity \cite{Dey2012}, however the precise dictionary is not known. 
The mirror-dual linear quivers have the same M-theory holographic backgrounds.
These M-theory backgrounds can be reduced to type IIA and T-dualized to type 
IIB in two ways by exchanging the M-theory and T-duality circles, 
leading to two IIB backgrounds in $S$-dual frames.

The backreacted geometry in the large~$N$ limit takes the form 
$AdS_4 \times S^7/(\bZ_{2\nu} \times \bD_{L})$, where the two orbifolds act separately on the 
two $S^3 \simeq SU(2)$ inside $S^7$. Recalling that the volume of $S^7$ is $\pi^4/3$ and the 
order of the $\bZ_{2\nu}$ and $\bD_{L}$ quotients are $2\nu$ and $4L$ 
respectively, we obtain the holographic prediction
\beq
C = \frac{6}{\pi^6} \frac{\pi^4}{24 \nu L} = \frac{1}{4 \pi^2 \nu L}\,,
\eeq
in perfect agreement with \eqref{DgenJmuresult}.
It would be interesting to find a similar nice geometrical interpretation in holography for 
the coefficient of subleading term
$B/\sqrt C$ in \eqref{freeEnergy} but this coefficient is very subtle and not even 
fully understood in the case of ABJM \cite{Bergman:2009zh}.

A localization calculation in $AdS_4$ \cite{Dabholkar:2014wpa} suggests that the result 
would be universal for any conformal 3d theory with an M-theory dual and enough 
supersymmetry. The $\hat D$-quivers (and their mirrors) studied in this paper fall into 
this class, and indeed we found the Airy function behavior, like for the $\hat A$-quivers. 
Note that this calculation does not predict the values of the $C$ and $B$ coefficients, 
rather they serve as arbitrary parameters of the calculation.

In our analysis of mirror symmetry, we left aside the $\hat D$-quivers whose naive mirror duals 
are bad linear quivers \cite{Dey2014a}. These are $\hat D$-quivers with 
numbers of fundamental hypermultiplets in a pair of terminating $U(N)$ nodes differing by 
two or more. The bad linear quivers have 
a divergent matrix model leading to a non trace class density operator. 
One example considered in \cite{Dey2014a} for instance, is the naive duality 
between a $\hat D_4$-quiver 
with two fundamental hypermultiplets on an external $U(N)$ node, and an $Sp(2N)^2$ linear quiver 
with four fundamental hypermultiplets on one $Sp(2N)$ node, and zero on the other. 
Manipulations as in \eqn{rhomanip} give the Wigner transformed density 
operator of the $\hat D_4$-quiver as
\beq
\frac{1}{\ch^2 p} \star \frac{\sh 2 p}{\ch^2 q} \star \frac{\sh 2 p}{\ch^6 p}\,.
\eeq
The Wigner transformed density operator of the naive mirror dual linear quiver is
\beq
\label{badmirrorwig}
\frac{1}{\ch p} \star \sh 2q \star \frac{1}{\ch p} \star \frac{\sh 2q}{\ch^8 q}\,.
\eeq
these operators would be related by the canonical transformation \eqn{mirrortrans} 
(and conjugation by $\frac{1}{\ch^2 p} $), if we also included the replacement 
(with $n=1$ and $m=0$)
\beq
\label{regreplace}
\frac{1}{\ch^n p} \star \frac{\sh 2q}{\ch^mp} \star \frac{1}{\ch^n p} 
\rightarrow \frac{\sh 2q}{\ch^{2n+m} p}
\eeq
This is
an {\it ad hoc} regularization of the partition function, transforming a non trace class density 
operator to one that is trace class, based on the assumption that the regularized partition 
function should satisfy the naive mirror symmetry. In fact, the replacement 
\eqn{regreplace} (with suitable $n$, $m$) 
easily regularizes and identifies a mirror dual for any bad linear quiver with 
no fundamental hypermultiplets on one or both terminal nodes, 
provided the theory has in total at least four fundamental hypermultiplets. 

It would be interesting to understand if 
\eqn{regreplace} can be derived from a proper regularization of
the divergent integrals in the matrix model.

One can think of several extensions to our work. Having found a Fermi gas formalism for the 
$\hat A$ and $\hat D$ quivers, it is natural to ask whether a Fermi gas formalism for the 
quiver theories of type $\hat E_{6,7,8}$ exist. 
To study these theories one needs to find suitable identities to put the corresponding matrix 
models in the form of a partition function for non-interacting particles in 1d.

Further generalizations involve necklace quivers with alternating $Sp$ and $SO$ groups. 
Likewise one can consider linear quivers with terminating $SO$ nodes or symmetric 
hypermultiplets (see \cite{Mezei2014}). One could also try to find generalizations of the 
$\hat D$-quivers, replacing one of the $\hat D$ ends with some other gauge groups or matter 
fields.

\subsection*{Acknowledgments} 
We would like to thank Cyril Closset, 
Stefano Cremonesi, Noppadol Mekareeya, Cristian Vergu and Sanefumi Moriyama 
for fruitful discussions.
N.D. would like to thank CERN (via the CERN-Korea Theory Collaboration), 
KIAS, Nagoya University and Humboldt University for hospitality during the course 
of this work. 
The research of N.D. is underwritten by an STFC advanced fellowship. The 
CERN visit was funded by the National Research Foundation (Korea).
The research of J.F. is funded by an STFC studentship ST/K502066/1.
B.A. acknowledges support by the ERC Starting Grant N. 304806, ``The Gauge/Gravity Duality and Geometry in String Theory.''.

\appendix

\section{Matrix models for quiver theories}
\label{sec:matrixmodels}

We provide in this appendix the factors appearing in the $\hat D$-quiver and linear 
quiver matrix models. Each factor is associated to an $\cN=4$ multiplet of the theory and 
is a function of the eigenvalues of the matrix model.

For $U(N)$ nodes we have the ingredients
\beq
\begin{tabular}{rll}
Eigenvalues:&$\lambda_i$\quad$i=1,\cdots,N$\,,
\\[1mm]
Weyl group order:&$|W|=N!$\,,
\\[1mm]
Vector multiplet:&$\displaystyle Z_{\text{vec}} = \prod_{i<j} \sh^2(\lambda_i- \lambda_j)$\,,
\\[1mm]
\parbox{2.8cm}{Fundamental hypermultiplet:}&
$\displaystyle Z^{\text{fund}}_{\text{hyper}} = \prod_{i} \frac{1}{\ch(\lambda_i+\mu)}$\,,
\\[1mm]
\parbox{2.8cm}{Antisymmetric hypermultiplet:}&
$\displaystyle Z^{\text{asym}}_{\text{hyper}} = \prod_{i<j} \frac{1}{\ch(\lambda_i + \lambda_j +M)}$\,.
\end{tabular}
\eeq
The expression for $U(2N)$ is the same with the obvious change $N\to 2N$, and where we use the convension of 
upper case $I,J$ for the indices.

For $Sp(2N)$ nodes the analogous expressions are
\beq
\!\!\!\begin{tabular}{rll}
Eigenvalues:&\!\!$\lambda_i$\quad$i=1,\cdots,N$\,,
\\[1mm]
Weyl group order:&\!\!$|W|=2^NN!$\,,
\\[1mm]
Vector multiplet:&\!\!$\displaystyle Z_{\text{vec}} 
= \prod_{i < j} \sh^2(\lambda_i - \lambda_j) \sh^2(\lambda_i + \lambda_j) \prod_i \sh^2(2\lambda_i)$\,,
\\[1mm]
\parbox{2.8cm}{Fundamental hypermultiplet:}
&\!\!$\displaystyle Z^{\text{fund}}_{\text{hyper}} = \prod_{i} \frac{1}{\ch(\lambda_i+\mu)\ch(\lambda_i-\mu)}$\,,
\\[1mm]
\parbox{2.8cm}{Antisymmetric hypermultiplet:}
&\!\!
$\displaystyle Z^{\text{asym}}_{\text{hyper}} 
=\prod_{i < j} \frac{1}{\ch(\lambda_i + \lambda_j +M)\ch(\lambda_i + \lambda_j-M)}
\prod_{i, j} \frac{1}{\ch(\lambda_i - \lambda_j+M)}$\,.
\end{tabular}\!\!
\eeq
Finally for the bifundamental hypermultiplets we have
\beq
\begin{tabular}{rll}
$U(N) \times U(M)$:&
$\displaystyle Z^{\text{bifund}}_{\text{hyper}} 
= \prod_{i=1}^N\prod_{j=1}^M\frac{1}{\ch(\lambda_i - \tilde \lambda_j - m)}$\,,
\\[1mm]
$Sp(2N) \times U(2N):$&
$\displaystyle Z^{\text{bifund}}_{\text{hyper}} 
= \prod_{i,J} \frac{1}{\ch(\lambda_i - \tilde \lambda_J - m)\ch(\lambda_i +\tilde \lambda_J+m)}$\,,
\\[1mm]
$Sp(2N) \times Sp(2N):$&
$\displaystyle Z^{\text{bifund}}_{\text{hyper}} 
=\prod_{i,j}\prod_\pm 
\frac{1}{\ch(\lambda_i+\tilde \lambda_j \pm m)\ch( \lambda_i-\tilde \lambda_j\pm m)}$\,.
\end{tabular}
\eeq

\section{Combinatorics of the permutations $R\tau^{-1}R \tau$ }
\label{sec:RRtauresults}

In this appendix we show how we can simplify the partition function \eqn{ZD4fermi} 
\beq
\label{ZD4fermiapp}
Z(N)= \frac{1}{2^{2N} N!^2 } \sum_{\tau \in S_{2N} } (-1)^{ \tau} \int d^N \lambda 
\prod_{k \in \cK(\tau)} (-1)^{s (k) + s(\tau (k)) }
\rho(\lambda_k, \lambda_{R \tau^{-1} R \tau (k)})\,,
\eeq
by studying more closely the composite permutation $R \tau^{-1} R \tau $ 
and the set $\cK(\tau)$ of $N$ integers in $1,\cdots, 2N$ such that 
$R(\cK(\tau))=\overline{\cK(\tau)}$ and $R \tau^{-1} R \tau(\cK(\tau))=\cK(\tau)$.

Let us label the $N$ integers in $\cK(\tau)$ by $k_1,\cdots, k_N$, such that the action
of $R \tau^{-1} R \tau $ on $\cK(\tau)$ can be represented in terms of a permutation 
$\sigma_\tau\in S_N$ as
\beq
R \tau^{-1} R \tau (k_i) = k_{\sigma_\tau (i) }\,.
\eeq
We can immediately see that $R \tau^{-1} R \tau $ acts on elements $R(k_i)$ in the 
compliment of $\cK(\tau)$ by the inverse permutation $\sigma^{-1}_\tau $
\beq
R \tau^{-1} R \tau ( R (k_i)) = R (R \tau^{-1} R \tau)^{-1} (k_i) = R (k_{\sigma^{-1}_\tau (i)} )\,.
\eeq
This property means that $R \tau^{-1} R \tau $ is composed of pairs of cycles that take the form%
\footnote{Indeed there is a freedom in choosing the set $\cK(\tau)$ by including the elements of
either one of each pair of cycles.}
\beq
\label{cycles}
(k_1 k_2 \cdots k_l)(R(k_l) R(k_{l-1}) \cdots R(k_1) )\,.
\eeq
For a given $\sigma\in S_N$ we can easily find $\tau\in S_{2N}$ such that $\sigma=\sigma_\tau$ by 
for example taking for each $l$-cycle in $\sigma$ the $2l$-cycle in $\tau$
\beq
(k_l R(k_l) k_{l-1} R(k_{l-1}) \cdots R(k_1) )\,.
\eeq
This particular 
choice of $\tau$ is useful because $\tau(k_i) = R(k_i)$, and it is made up of only cycles of even length, 
so we can easily compute
\beq
\label{minusfactor}
(-1)^\tau \prod_{k \in \cK(\tau)} (-1)^{s (k) + s(\tau (k))} = (-1)^{n_\tau} (-1)^N
= (-1)^{n_{\sigma_\tau}} (-1)^N = (-1)^{\sigma_\tau}\,,
\eeq
where we recall that
\beq
s(k) = \begin{cases} 0\,, &k=1,\cdots, N\,,\\
1\,, & k =N+1,\cdots, 2N\,, \end{cases}
\eeq
and $n_\tau$ counts the number of cycles in $\tau$. The second equality in \eqn{minusfactor} follows 
because each cycle in $\tau$ gives rise to a cycle in $\sigma_\tau$, and in the third equality 
we recognised that the expression appearing is nothing but the signature of $\sigma_\tau$.

Of course, there are many possible choices of $\tau$ giving rise to the same 
$R \tau^{-1} R \tau$ and we should check that all of them reproduce \eqn{minusfactor}. 
From any particular $\tau$ we can generate all $\tau$ giving rise to the same 
$R \tau^{-1} R \tau$ by taking
\beq
\label{tautransform}
\tau \rightarrow \pi \tau\,, 
\qquad 
\pi^{-1} R \pi = R
\eeq
To solve this condition, $\pi$ can be any permutation of the form $\pi = \pi_1 \pi_2$, 
where $\pi_1$ is any combination of the two cycles appearing in $R$ ($2^N$ possibilities), 
and $\pi_2$ acts as two copies of some $S_N$ permutation ($N!$ possibilities)
\beq
\pi_2 (i) = \sigma' (i), \qquad \pi_2 (N+i) = N + \sigma' (i), 
\qquad i = 1,\cdots, N, \quad \sigma' \in S_N\,.
\eeq
It is easy to check that any deformation \eqn{tautransform} leaves the left-hand side of 
\eqn{minusfactor} invariant, and so the right-hand side indeed holds for any $\tau$. 
With this simplification we can rewrite the partition function \eqn{ZD4fermiapp} as
\beq
\label{ZNhalfsimp}
Z(N)= \frac{1}{2^{2N} N!^2 } \sum_{\tau \in S_{2N} } (-1)^{ \sigma_\tau} \int d^N \lambda 
\prod_{i=1}^N \rho(\lambda_{k_i}, \lambda_{k_{\sigma_\tau(i)} })\,,
\eeq
The summand of \eqn{ZNhalfsimp} depends only on the conjugacy class 
of $\sigma_\tau$, determined by the number of cycles $m_l$ of length $l$. 
This means we can convert the sum over $S_{2N}$ permutations to a sum over conjugacy classes
of $S_N$, if we know the combinatorics of the map from $\tau$ into the conjugacy class of
$\sigma_\tau$.

We have already counted how many $\tau$ give rise to any given 
$R\tau^{-1} R \tau$ permutation ($2^N N!$), so we just need to compute how many distinct 
$R\tau^{-1} R \tau$ 
permutations are associated with a $\sigma_\tau$ of a given conjugacy class. We know that 
$R\tau^{-1}R \tau$ generates all possible permutations made up of pairs of cycles as in \eqn{cycles}, 
and we should count how many of them have $m_l$ pairs of cycles of length $l$.
Suppose we fix how we assign to $k_1,\cdots, k_N$ and $R(k_1),\cdots, R(k_N)$ the integers 
$1,\cdots, 2N$. 
With this restriction we are just left with counting the number of ways to distribute the $k_i$ among the 
cycles, which is the same as counting the number of $S_N$ permutations in a given conjugacy class
\beq
\label{nml}
\frac{N!}{\prod_{l=1}^N l^{m_l} m_l!}\,. 
\eeq
We then have the added possibility of generating more distinct $R\tau^{-1}R \tau$ 
(without altering the conjugacy class) by exchanging $k_i \leftrightarrow R(k_i)$. 
Note however that simultaneously exchanging all of the $k_i$ within a given cycle leaves 
$R\tau^{-1}R \tau$ invariant, so this generates just an additional factor of
\beq
\frac{2^N}{2^{n_{\sigma_\tau}}}\,,
\eeq
Putting all this together and relabelling 
$\lambda_{k_i} \rightarrow \lambda_i$, \eqn{ZNhalfsimp} becomes
\beq
\label{ZNgeneral-app}
Z(N)= \frac{1}{ N! } \sum_{\sigma \in S_{N} } \frac{(-1)^\sigma}{2^{n_\sigma}} \int d^N \lambda 
\prod_{i=1}^N \rho(\lambda_i, \lambda_{\sigma(i)})\,,
\eeq
where we have absorbed the factor \eqn{nml} to promote the sum over $S_N$ conjugacy classes to a 
sum over $S_N$ permutations.

\section{Degeneracy of the spectrum}
\label{sec:degenspectrum}

In this appendix we show that for vanishing masses and FI parameters 
the expressions for the partition functions found in the main text 
in terms of density operators \eqref{ZNgeneral} (which is identical to \eqn{ZNgeneral-app} above), 
correspond to the partition function of fermions on a semi-infinite line \eqref{ZNprojplus}. We do this 
by proving that the spectrum of $\rho$ splits into odd and even states whose spectrum is identical.%
\footnote{Apart for a single even zero-mode, which is non-normalizable.}

The first statement amounts to $\rho$ commuting with the reflection operator $\hat R$ which acts on 
states by $\hat R \ket{\lambda} = \ket{-\lambda}$. 
For vanishing masses and FI parameters the density operator $\rho$ in \eqref{rhoDgen} is given by 
a sequence of even or odd functions of $p$ or $q$, with precisely two odd functions. 
Since $\hat R f(q) = f(- q) \hat R$, $\hat R f(p) = f(-p) \hat R$, for any function $f$, 
we see that $\hat R$ commutes with $\rho$.

To show that the spectrum of odd and even states 
is the same, we prove that $\Tr(\rho^l \hat R)=0$ for all $l$.
Let us focus first on the case $l=1$.
We notice that the density operator takes the specific form
\beq
\rho= B^{(0)}_D (p,q) \,\tilde\rho \,B^{(L)}_D (p,q) \,\tilde\rho^\dagger\,,
\eeq
with $\tilde\rho$ a sequence of even functions of $p$ or $q$, $\tilde\rho^\dagger$ is the 
Hermitean conjugate and
\beq
B^{(a)}_D (p,q)= \left( F^{(a)}( q) \frac{\sh p }{\ch p} F^{\prime(a)}( q) 
+F^{\prime(a)}( q) \frac{\sh p }{\ch p} F^{ (a)}( q) \right), 
\qquad a = 0, L\,.
\eeq
Using the properties%
\footnote{Here $\dagger$ acts as transposition, since we consider only real operators.}
\beq
\hat R^\dagger = \hat R \, ,
\quad 
f(q)^\dagger = f(q)\,, 
\quad 
f(p)^\dagger = f(-p)\,,
\eeq
we can derive the chain of equalities
\bal
\Tr (\rho \hat R) &= \Tr \left( (\rho\hat R)^\dagger \right) = \Tr \left( \hat R \rho^\dagger \right)
= \Tr \left( \hat R \,\tilde\rho \,B^{(L)}_D (p,q) \,\tilde\rho^\dagger \,B^{(0)}_D (p,q) \right)
\\&= - \Tr \left( B^{(0)}_D (p,q) \,\tilde\rho \,B^{(L)}_D (p,q) \,\tilde\rho^\dagger \,\hat R \right)
= - \Tr (\rho \hat R)\,,
\eal
where we have used $(B^{(a)}_D (p,q))^\dagger = -B^{(a)}_D (p,q) $, the cyclicity of the trace and commuted 
$\hat R$ and $ B^{(0)}_D (p,q) $, producing a minus sign.
This yieds $\Tr (\rho \hat R) = 0$. The argument generalizes easily $l \ge 2$.

To derive \eqref{ZNgeneral}, notice that the effect of the projection 
$\frac{1\pm\hat R}{2}$ in \eqref{ZNprojplus}, \eqref{ZNprojminus} is to add a factor of $1/2$ 
to every cycle in a given permutation $\sigma \in S_N$,
\bal
\int \prod_{k=1}^l d\lambda_{i_k} \, \bra{\lambda_{i_l}} \rho \, \bigg( \frac{1\pm\hat R}{2} \bigg) \ket{\lambda_{i_1}} 
\prod_{k=1}^{l-1} \bra{\lambda_{i_k}} \rho \, \bigg( \frac{1\pm\hat R}{2} \bigg) \ket{\lambda_{i_{k+1}}} 
&= \Tr \left( \bigg( \rho \, \frac{1\pm\hat R}{2} \bigg)^l \right) 
\\
= \Tr \bigg( \rho^l \, \frac{1\pm\hat R}{2} \bigg)
&= \frac 12 \, \Tr \left( \rho^l \right).
\eal

The same results hold for density operators of linear quivers \eqref{rhoSpAgen} at vanishing 
masses and FI parameters. The arguments are the same except that one must consider the 
momentum space basis $\ket{p}$, with for instance $\Tr \hat A = \int dp \bra{p} \hat A \ket{p}$.

\section{Relation with previous Fermi gas formulation of $Sp(2N)$ quivers}
\label{sec:prevfermi}

In this appendix we compare our results to those in \cite{Mezei2014}, which also developed a 
Fermi gas approach to $Sp$ quivers. As we show below, the two formulations are equivalent, but in the 
final expression for the coefficients $B$ in the Airy function, we disagree with their results and explain why.

The overlap between theories discussed in \cite{Mezei2014}
and our work is the single node $Sp$ quiver with an antisymmetric
hypermultiplet and $n$ fundamental hypermultiplets. The partition function is given by the matrix model
\beq
\label{ZN1sp}
Z(N)= \frac{1}{4^N N!} \int d^N \lambda \prod_{i=1}^N \frac{\sh^2 2 \lambda_i \ch 2 \lambda_i }{ \ch^{2n} \lambda_i } 
\frac{ \prod_{i<j} \sh^2 (\lambda_{i} - \lambda_{j} ) \sh^2 (\lambda_{i} + \lambda_{j} ) }
{ \prod_{i,j}\ch (\lambda_{i} - \lambda_{j} ) \ch (\lambda_{i} + \lambda_{j} ) }\,.
\eeq

In \cite{Mezei2014} the matrix model \eqn{ZN1sp} was manipulated with a modified Cauchy identity
\beq
\label{modcauchy}
\frac{\prod_{i<j} (x_i- x_j) (y_i- y_j) (x_i x_j -1) (y_i y_j -1)}{\prod_{i,j} (x_i + y_j) (x_i y_j +1) } = 
\sum_\sigma (-1)^\sigma \prod_i \frac{1}{(x_i + y_{\sigma(i)} ) (x_i y_{\sigma(i)} +1)}\,.
\eeq
This identity, with the usual the usual replacements $x_i \rightarrow e^{\lambda_i } $, 
$y_i \rightarrow e^{\lambda_i}$, gives
\beq
Z(N)= \frac{1}{ 4^N N!}\sum_{\sigma \in S_N} (-1)^\sigma \int d^N \lambda \prod_{i=1}^N 
\frac{\prod_i \sh^2 2 \lambda_{i} \ch 2 \lambda_{i}}{ \prod_i \ch^{2n} \lambda_{i} }
\frac{1 }{ \ch (\lambda_{i} - \lambda_{\sigma(i)} ) \ch (\lambda_{i} + \lambda_{\sigma(i) } ) } 
\eeq

Using the relations%
\footnote{Note that $\hat R$ commutes with $\frac{1}{\ch p}$, and so we can write 
$\frac{1+\hat R}{2\ch p}$ ( $= \frac{1}{\ch p} \frac{1 + \hat R}{2}$ ) which would otherwise be ill defined} 
\bal
\label{Rid}
\frac{\ch \pi \lambda_1 \ch \lambda_2 }{\ch (\lambda_1-\lambda_2) \ch (\lambda_1+\lambda_2) } 
&= \bra{ \lambda_1 } \frac{1+\hat{R}}{2 \ch p } \ket{\lambda_2}\,, \\
\frac{\sh \lambda_1 \sh \lambda_2 }{\ch (\lambda_1-\lambda_2) \ch (\lambda_1+\lambda_2) } 
&=\bra{ \lambda_1} \frac{1-\hat{R} }{2 \ch p } \ket{ \lambda_2 }\,,
\eal
from which one immediately has the corollary 
\beq
\label{projident}
\frac{1-\hat{R} }{2 \ch p } = \frac{\sh q}{\ch q} \frac{1+\hat{R}}{2 \ch p }\frac{\sh q}{\ch q}\,,
\eeq
they rewrite the partition function in two equivalent forms
\beq
\label{ZNprojpm}
Z(N)= \frac{1}{N!}\sum_{\sigma \in S_N} (-1)^\sigma 
\int \prod_{i=1}^N d\lambda_i \,\prod_{i=1}^N 
\bra{\lambda_i} \rho_{\pm} \bigg( \frac{1 \pm \hat R}{2} \bigg) \ket{\lambda_{\sigma(i)}}\,,
\eeq
where
\bal
\label{rho1spproj}
\rho_+ &= \frac{\sh^2 q \ch 2 q}{\ch^{2n } q}\frac{1}{\ch p }\,,\\
\rho_- &= \frac{\ch 2 q}{\ch^{2n-2} q} \frac{1}{\ch p } \, .
\eal

Indeed, this looks very similar to our rewriting of the matrix model as 
\eqn{ZNprojplus} or \eqn{ZNprojminus}
\beq
\label{ZNprojpm2}
Z(N)= \frac{1}{N!}\sum_{\sigma \in S_N} (-1)^\sigma 
\int \prod_{i=1}^N d\lambda_i \,\prod_{i=1}^N 
\bra{\lambda_i} \rho \bigg( \frac{1 \pm \hat R}{2} \bigg) \ket{\lambda_{\sigma(i)}}\,,
\eeq
but with a different density operator \eqn{rho1sp}
\beq
\label{rho1spapp} 
\rho = \frac{1}{2} \frac{\sh 2 q}{\ch^{2 n} q } 
\left( \sh q \frac{1}{\ch p} \ch q + \ch q \frac{1}{\ch p} \sh q \right).
\eeq

All those expressions are in fact equivalent, since we can show that the spectrum of even states of 
$\rho$ agrees with that of $\rho_+$ and the spectrum of odd states to that of $\rho_-$.
Indeed the projected operators are similar to each other by the simple manipulations 
\bal
\label{oddproj}
\rho \, \frac{1 - \hat{R} }{2 } &= \frac{1}{2} \frac{ \sh 2q }{\ch^{2n} q} 
\bigg( \sh q \frac{1 - \hat{R}}{2 \ch p} \ch q
+ \ch q \frac{1 + \hat{R}}{2 \ch p} \sh q \bigg) 
\\&= \frac{1}{2} \frac{ \sh 2q }{\ch^{2n} q} \bigg( \sh q \frac{1-\hat{R}}{2 \ch p} \ch q
+ \frac{\ch^2 q}{\sh q} \frac{1-\hat{R}}{2 \ch p} \ch q \bigg)
\\&= \frac{ \ch 2q \ch q }{\ch^{2n} q} \frac{1-\hat{R}}{2 \ch p} \ch q 
=\frac{1}{\ch q} \bigg( \rho_{-} \frac{1- \hat R}{2}\bigg) \ch q\,,
\eal
where in the first line we used that $\hat R$ commutes with even and anticommutes with 
odd functions of $q$, and in the second line we used \eqn{projident}. Similar manipulations 
lead to the relations
\bal
\frac{1}{\ch q} \bigg( \rho_{-} \frac{1- \hat R}{2}\bigg) \ch q 
= \frac{1}{\sh q} \bigg( \rho_{+} \frac{1+ \hat R}{2}\bigg) \sh q 
=\frac{\ch q}{\ch 2q \sh q} \bigg(\rho \, \frac{1 + \hat{R} }{2 }\bigg) \frac{\ch 2q \sh q}{\ch q}\,.
\eal

We can also derive the Airy function expression for these theories based on our analysis in 
sections~\ref{sec:mirror} and~\ref{sec:partition}. The mirrors of this class of theories are 
$\hat D$ quivers with $(n-1)$ $U(2N)$ nodes and and a single fundamental hypermultiplet 
on one $U(N)$
node. The asymptotic expansion of the grand potential 
\eqn{DgenJmuresult} for this theory ($L=n-1$, $\nu=\Delta=\frac{1}{2}$, 
$\Sigma_1 = \Sigma_2 =0$) is given by 
\beq
C= \frac{1}{2 \pi^2 (n-1)}\,,
\qquad
B= \frac{1}{8} \left( -n -2 + \frac{1}{n-1} \right).
\eeq

The coefficient $C$ is the same as found in \cite{Mezei2014}, but $B$ is not. The reason for 
the discrepancy is that in our formulation, the operator $\rho$ has degenerate odd/even spectrum, 
as shown in appendix~\ref{sec:degenspectrum}. While the odd and even spectra agree with 
$\rho_+$ and $\rho_-$ of \cite{Mezei2014}, those latter operators do not have degenerate 
spectra. The saddle point calculation of \cite{Mezei2014} based on the Fermi surface of $\rho_-$ 
considered the full spectrum of this operator and divided the end result by 2. Since the 
operator does not have a degenerate spectrum, this is merely an approximation that is good 
enough to evaluate the leading order term $C$, but fails for the subleading coefficient $B$.

\section{Truncation of the $\epsilon$ expansion}
\label{sec:epsilontruncation} 

Here we show that the $\epsilon$ expansions of $C$ and $B$ coefficients in \eqn{Jmuasymp} 
truncates at order $\epsilon^0$ and $\epsilon^2$ respectively, 
adapting a similar proof from \cite{Marino2012} for $\hat A$ quiver theories.

We recall that corrections to $C$, $B$ and $A$ coefficients at order $\epsilon^n$ 
can be computed from a single residue involving $Z_{l(n)}$
\bal
J(\mu) = \sum_{n\geq 0} \epsilon^n J_{(n)}(\mu), \qquad J_{(n)}(\mu) = 
-\frac{1}{2} \mathop{\text{Res}}\limits_{l=0} \Gamma(l)\Gamma(-l) 
Z_{l(n)} e^{l \mu} +\cO (e^{- \alpha \mu } ), \qquad \alpha >0\,.
\eal
From this expression it is clear that if $Z_{l(n)}$ vanishes at $l=0$,
then $J_{(n)}(\mu)$ can only correct the $A$ coefficient. We should prove then that $Z_{l(n)}$ has at least a 
simple zero at $l=0$ for all $n>2$. 

It is useful to consider the Wigner-Kirkwood expansion of $Z_l$ \cite{Marino2012}. The idea is to
express $Z_l$ in terms of the Fermi gas Hamiltonian, $H_W = - \log_\star \rho_W$
\bal
\label{Zlkirkwood0}
Z_l = \sum_{r \geq 0} \frac{(-l)^r}{r!} \int \! dp dq \, e^{- l H_W} \cG_r, \qquad \cG_r = \big( ( \hat H - H_W)^r \big)_W\,.
\eal 
We recall that the $\hat D$ density operator is given by \eqn{rhoDgenwig}
\bal
\label{rhoDgenwigapp}
\rho_W = \sqrt[\leftroot{-2}\uproot{2}\star]{e^{S (p) +2 U_\lambda (q)}} \star
e^{T(p)} &\star 
\bigg( \prod_{k=1}^{\lambda-1}\mathstrut{\vphantom\prod}_\star e^{U_{\lambda-k} (q) } 
\star e^{T(p)}\bigg)\star e^{S (p) +2 U_0 (q) }
\\&\star \bigg( \prod_{k=1}^{\lambda-1}\mathstrut{\vphantom\prod}_\star 
e^{T(p)} \star e^{U_{k} (q) } \bigg) \star e^{T(p)} 
\star\sqrt[\leftroot{-2}\uproot{2}\star]{e^{S (p) +2 U_\lambda (q)}} \, .
\eal
Since \eqn{rhoDgenwigapp} has an expansion in purely even powers of $\epsilon$, 
$H_W$ likewise takes the form
\bal
H_W = H_{(0)} + \epsilon^2 H_{(2 )} + \epsilon^4 H_{(4 )} + \cdots\,.
\eal
This gives 
\bal
\label{Zlkirkwood}
Z_l = \sum_{r \geq 0} \frac{(-l)^r}{r!} \int \! dp dq \, e^{- l H_{(0)} }
\left(1 -l \sum_{n \geq 1} \epsilon^{2n}H_{(2n) } + \frac{l^2}{2} 
\bigg( \sum_{n \geq 1} \epsilon^{2n}H_{(2n) } \bigg)^2 + \cO (l^3) \right) \cG_r\,.
\eal 
We know that all of the integrals that appear in this expansion are of the form 
\eqn{trigint}
\beq 
\label{trigintapp}
\int dx \frac{\thh^{a} x }{\ch^b x } = \frac{ (1 + e^{i \pi a} ) 
\Gamma\left( \frac{a +1}{2}\right)\Gamma\left( \frac{b}{2}\right)}
{2^{b+1} \pi \Gamma\left( \frac{a+b+1}{2}\right)}\,.
\eeq 
The coefficients $a$ and $b$ are given by linear functions of $l$ and so the integrals over 
$p$ and $q$ in \eqn{Zlkirkwood} can each produce at most a simple pole at $l=0$, 
from the $\Gamma\left( \frac{b}{2}\right)$. 
Therefore, we need only concern ourselves with 
terms in \eqn{Zlkirkwood} with prefactors of at most order $l^2$. 
Discarding also (most of) the terms of order 
$\epsilon^2$ or less, we are left with%
\footnote{The final term $(H_W \star H_W - H_W^2)$ has still some 
$\epsilon^2$ piece, but as we shall see this also has 
at least a simple zero at $l=0$ } 
\bal 
\label{remainingterms} 
&\int \! dp dq \, e^{- l H_{(0)} } \left(-l \sum_{n \geq 2} \epsilon^{2n}H_{(2n) }+ \frac{l^2}{2} 
\bigg( \sum_{n \geq 1} \epsilon^{2n}H_{(2n) } \bigg)^2\right) 
\\& \hspace{5cm} + \frac{l^2}{2} \int \! dp dq e^{- l H_{(0)} } (H_W \star H_W - H_W^2)\,,
\eal
For simplicity let us now restrict the density operator \eqn{rhoDgenwigapp} 
to cases with $U_0 (q) = U_\lambda (q) = 0$.%
\footnote{This corresponds to restricting the $\hat D$ quiver theories shown 
in figure~\ref{fig:Dgen} to a subclass with 
$n^{(0)} =n^{\prime (0)}$ and $n^{(L)} =n^{\prime (L)}$. 
We expect that theories outside this subclass 
also have $B$ and $C$ coefficients truncating at $\epsilon^2$. 
We have verified that this holds true 
up to $\epsilon^4$.}
With this restriction, the exponentials in \eqn{rhoDgenwigapp} can be freely 
exchanged with star exponentials, and we
can straightforwardly evaluate the star logarithm by the star product version 
of the Baker-Campbell-Hausdorff expansion. The leading term gives
\beq
\label{elH0}
e^{- l H_{(0)} } = \left( \frac{\thh^{2} p}{\ch^{2 L} p \ch^{2 \nu} q } \right)^l\,.
\eeq
All $\epsilon$ corrections to $H_W$ are then given by nested star commutators 
involving $T(p)$, $S(p)$ and $U_k(q)$. An example of such a term contributing to $H_{(4)}$ would be
\beq
\label{nestcomexample}
[ T(p), [ T(p), [T(p), [T(p), U_k(q) ]]]]_\star 
= \frac{\epsilon^4}{16 \pi^4} \dot{T}^4 U_k^{(4)} + \cO (\epsilon^6)\,,
\eeq
where we have used
\beq
{} [ f, g ]_\star = f \star g - g \star f = i \frac{\epsilon}{2\pi} \{ f, g \} + \text{higher derivative terms}\,.
\eeq
As illustrated in \eqn{nestcomexample}, in order to have a term at order $\epsilon^n$ with 
only single derivatives acting on functions of $p$ (or $q$), such a term has a single 
function of $q$ (or $p$) with an $n$\textsuperscript{th} derivative. 

This is important, because every term in the epsilon expansion has therefore 
at least one multiple derivative of $S$, $T$ or $U_k$. 
From \eqn{UTdefs} it follows that these derivatives take the form
\beq
U_k^{(n)}(q) 
= \frac{1}{\ch^2 q} \sum_{a \in \mathbb{Z}} \sum_{b\geq 0} C_{a b} \frac{\thh^{a} q}{\ch^{b} q}\,,
\qquad
n\geq2\,,
\eeq
and similarly for $S$, $T$ with $q\rightarrow p$.

We now should combine these derivative terms with \eqn{elH0} in \eqn{remainingterms}, 
and integrate with \eqn{trigintapp}. It is clear that since the derivative terms contribute 
always a $\frac{1}{\ch^2 q}$ or $\frac{1}{\ch^2 p}$, the resulting Gamma 
functions can contribute at most a simple pole at $l=0$. This guarantees the terms in 
\eqn{remainingterms} with an $l^2$ outside have at least a simple zero, leaving us with
\beq
\label{remainingterms2}
- l \int \! dp dq \, e^{- l H_{(0)} } \sum_{n \geq 2} \epsilon^{2n} H_{(2n) }\,.
\eeq
By the same reasoning, terms with multiple derivatives on 
both a function of $p$ and a function of $q$ have an overall 
$\frac{1}{\ch^2 q} \frac{1}{\ch^2 p}$ which kills both of the poles one could get from integrating. 
The remaining terms in \eqn{remainingterms2} which don't obviously have a simple zero 
are those where all derivatives on functions of $p$ (or $q$) are first order, like \eqn{nestcomexample}. 
But such terms can be integrated by parts, for instance \eqn{nestcomexample} would give 
\beq
- l \int \! dp dq \, e^{- l H_{(0)} } \frac{\epsilon^4}{16 \pi^4} \dot{T}^4 U_k^{(4)} 
= l^2 \int \! dp dq \, e^{- l H_{(0)} } \frac{\epsilon^4}{16 \pi^4} \dot{T}^4 U_k^{(3)}H_{(0)}'\,.
\eeq
Integrating by parts pulls down an additional factor of $l$, which guarantees that there is 
an overall simple zero at $l=0$, since the integral on the right hand side still produces just a 
simple pole. Since we have shown that at order $\epsilon^4$ and higher $Z_l$ has 
at least a simple zero at $l=0$, this concludes the proof that $C$ and $B$ do not get contributions 
beyond order $\epsilon^2$.

\bibliographystyle{utphys2}

\bibliography{References}

\end{document}